\documentclass[12pt,tightenlines,eqsecnum,floats,showpacs,nofootinbib, aps, prd]{revtex4-2}
\usepackage{enumitem}
\usepackage{graphicx}
\usepackage{subcaption}
\usepackage{amsmath,amsthm}
\usepackage{amsfonts}
\usepackage{amssymb}
\usepackage{mathrsfs}
\usepackage[dvipsnames]{xcolor}
\usepackage{color}
\usepackage{epstopdf}
\usepackage[utf8]{inputenc}
\usepackage[normalem]{ulem} 
\usepackage{natbib}
\usepackage[colorlinks,urlcolor=NavyBlue,citecolor=NavyBlue,linkcolor=NavyBlue,pdfusetitle]{hyperref}
\usepackage[all]{hypcap}
\usepackage{orcidlink}
\usepackage{mathtools}

\usepackage{wasysym}

\theoremstyle{definition}

\theoremstyle{remark}
\newtheorem*{remark}{Remark}
\newtheorem*{remarks}{Remarks}

\newcommand{\scri}{\mathscr{I}}
\newcommand{\scrip}{\scri^{+}}
\newcommand{\scrim}{\scri^{-}}
\newcommand{\scripm}{\scri^{\pm}}
\newcommand{\inot}{i^{\circ}}
\renewcommand{\th}{\textrm{\thorn}}

\usepackage{CJKutf8}
\newcommand{\Ro}{\textrm{\begin{CJK}{UTF8}{min}ロ\end{CJK}}}
\newcommand{\Roo}{\Ro \kern-2.2ex{\raisebox{1.35ex}{ \scalebox{0.45}{$\circ$}}}\kern0.9ex{}}
\begin{document}
\title{From spatial to null infinity: Connecting initial data to peeling}
\author{Berend Schneider\orcidlink{0009-0004-4099-2574}}
\email{berend@uoguelph.ca}
\affiliation{Department of Physics, University of Guelph, Guelph, Ontario, Canada N1G 2W1}
\author{Neev Khera\orcidlink{0000-0003-3515-2859} }
\email{nkhera@tsinghua.edu.cn}
\affiliation{Department of Astronomy, Tsinghua University, Beijing 100084, China}
\affiliation{Department of Physics, University of Guelph, Guelph, Ontario, Canada N1G 2W1}
\begin{abstract}
The asymptotic structure of space-time is studied by imposing conditions on the asymptotics of the metric. These conditions are weak enough to include large classes of physically relevant isolated space-times, but have a rich enough structure to be able to define important physically meaningful quantities like mass, angular momentum, and gravitational waves. 

By using a unified expansion of the metric in a neighborhood of spatial infinity that includes a piece of null infinity, we connect the asymptotic expansions of solutions to Einstein's equations in the different asymptotic regimes. Within the class of space-times under consideration, we find a connection between the peeling properties of the Weyl scalars and symmetries of initial data near spatial infinity. In particular, we show that for initial data that to leading order is symmetric under parity + time reversal, $\Psi_2$ has the usual $1/r^3$ fall-off rate at null infinity. If, in addition, the subleading part of the data is antisymmetric under parity + time reversal, then $\Psi_1$ has the usual $1/r^4$ fall-off rate at future null infinity.

\end{abstract}

\maketitle

\section{Introduction}
In general relativity, the role of space-time fundamentally changes from the background to the foreground. As a consequence it is impossible to study isolated objects, such as black holes or compact binary coalescences, in isolation; the dynamics of the entire space-time is relevant. Thus, the metric on the entire space-time must be included for a consistent analysis of the physical system. Then, to have a complete description of the space-time it is necessary to have asymptotic conditions on the metric, which provide the boundary conditions on the space-time.

For the study of isolated systems the asymptotic limit of the metric can be expected to be relatively simple. In particular, for a vanishing cosmological constant, the metric can be expected to asymptotically approach a Minkowski metric. This asymptotic limit can be taken in three different ways: along future directed null directions to future null infinity $\scrip$, along past directed null directions $\scrim$, or along spatial directions to spatial infinity $\inot$.

In the seminal work of Bondi et al.~\cite{bondi}, the asymptotic conditions  for null infinity was treated in a fully nonlinear setting. In particular, this led to a surprising realization that the structure at null infinity is much richer than that of Minkowski space, containing an infinite dimensional enlargement of the symmetry group~\cite{sachs, sachs2}. This was later recast in a geometric language by Penrose ~\cite{rp}, by using a conformal completion to add null infinity as a boundary to the space-time manifold. An important consequence of their asymptotic conditions is the so-called \emph{peeling theorem}, which states that the Weyl scalars $\Psi_k$, where $0\leq k \leq 4$ have the following asymptotic fall-off rates at $\scrip$:
\begin{equation}
  \label{eq:Psi_i_peel_scrip}
  \Psi_k(u, r, \theta, \phi) = \frac{\Psi_k^\circ(u, \theta,\phi)}{r^{5-k}} + o\left(\frac{1}{r^{5-k}}\right)\,.
\end{equation}
Here $\Psi_4^\circ$ is the component of the field that describes the outgoing radiation, $\Psi_2^\circ$ contains information about the Bondi mass of the space-time, and $\Psi_1^\circ$ contains information about the angular momentum of the space-time.
Similarly, for asymptotically flat space-times at $\scrim$, the Weyl scalars have the fall-off rates
\begin{equation}
  \label{eq:Psi_i_peel_scrim}
  \Psi_k(v, r, \theta, \phi) = \frac{\Psi_k^\circ(v, \theta,\phi)}{r^{k+1}} + o\left(\frac{1}{r^{k+1}}\right)\,,
\end{equation}
where now $\Psi_0^\circ$ describes the ingoing radiation, and $\Psi_3^\circ$ contains information about the angular momentum.

While the peeling theorem seems reasonable, the question of whether or not it applies to physically relevant systems remains a delicate issue. It has been argued by many that the conditions imposed on $\scripm$ which lead to peeling are too strong. In particular, in~\cite{dcsk} for the class of perturbations considered it was shown that $\Psi_1$ and $\Psi_0$ do not satisfy the peeling rates~\eqref{eq:Psi_i_peel_scrip} at $\scrip$. See also~\cite{bieri} and the excellent series of papers `The Case Against Smooth Null Infinity' by Kehrberger~\cite{Kehrberger:2021uvf,Kehrberger:2021vhp,Kehrberger:2021azo,Kehrberger:2024clh,Kehrberger:2024aak} for more analyses and examples of space-times where peeling does not hold. Such violations have physically relevant consequences, because if $\Psi_1$ does not peel, it has implications on physical quantities like angular momentum and for techniques like CCE~\cite{Bishop:1996gt, Reisswig:2009rx, Moxon:2021gbv, Iozzo:2021vnq} and CCM~\cite{Ma:2024hzq} used to extract gravitational waves from numerical simulations.

Certainly, for generic initial data, peeling will not hold. Nonetheless, physical systems of interest such as binary black hole coalescences or scattering are not generic, but instead have several special characteristics. For instance, these space-times typically do not contain ingoing radiation at $\scrim$. Therefore, the question remains whether there are any astrophysically relevant space-times where assuming peeling is too strong. To address this question, we characterize the class of space-times for which peeling does hold. Furthermore, we will discuss its implications on the class of space-times considered in compact binary coalescences. As we will discuss in \S\ref{sec:discussion}, the results in this paper need to be extended to include scattering space-times, but could be used to analyse PN space-times as well as typical numerical simulations.

We perform the characterization by solving Einstein's equations in a neighborhood of spatial infinity, and study these solutions near null infinity. We do so by using a unified asymptotic framework that includes three regimes: the region near spatial infinity, the far past of future null infinity, and the far future of past null infinity. There are no conditions imposed on the metric in the limit to null infinity; Instead, the asymptotics at null infinity are determined from the solutions to Einstein's equations. In particular, this will relate the asymptotics of the fields at null infinity directly to the initial data. Additionally, it relates the data at past null infinity to future null infinity. We will see that the peeling property of the space-time is intimately tied to the parity plus time reversal ($\mathcal{PT}$) symmetry, at different orders in the expansion of the metric.

To evolve the space-time from an initial slice, we use an asymptotically flat form of the metric at spatial infinity and evolve it using Einstein's equations. The asymptotic conditions at spatial infinity were first treated by using a 3+1 decomposition by ADM in~\cite{adm}. This was set into a geometrical setting by adding a boundary point $\inot$ in~\cite{rg-jmp}. A four-dimensional geometric formulation was introduced in~\cite{Ashtekar:1978zz}. A coordinate based asymptotic expansion in four dimensions was later done in~\cite{rbbs}. See also \cite{am3+1, aajr, henneaux2, henneaux3} for other approaches.

Traditionally, coordinate based approaches to $\scrip$, $\scrim$ and $\inot$ use distinct incompatible coordinate systems. For $\scrip$, Bondi coordinates $(u,r, \theta, \phi)$ are used, where $u$ is a retarded time coordinate, and $\scrip$ is the $r\to\infty$ limit. The metric near $\scrip$ is therefore expressed as an expansion in $1/r$. This expansion typically \emph{assumes} Peeling to hold \textit{a priori}. At past null infinity the story is similar, except the coordinates $(v,r, \theta, \phi)$ are used, where $v$ is an advanced time coordinate. At $\inot$ on the other hand, a hyperboloidal coordinate $\rho = \sqrt{r^2-t^2}$ is used~\cite{Ashtekar:1978zz, rbbs}. Spatial infinity $\inot$ is expressed as the $\rho\to\infty$ limit, and the metric near $\inot$ is expressed as an expansion in $1/\rho$. In \cite{Compere:2023qoa}, a matching procedure between these different expansions is described. This is challenging because infinitely many terms in $1/\rho$ at $\inot$ contribute to a single order in $1/r$ at $\scrip$ and $\scrim$. Similarly, infinitely many orders in $1/r$ at $\scrip$ and $\scrim$ contribute to a single order in $1/\rho$ at $\inot$. Thus, an assumption on the form of the metric needs to be made to perform the matching. In particular, this includes an assumption on the level of peeling at null infinity. Thus, this method is not suited for the the problem we seek to address in this paper.

Instead, we would like to consider initial data of Einstein's equations near spatial infinity and evolve them towards null infinity to find conditions for peeling \emph{in terms of the initial data}. Because we are in an asymptotically flat space-time, we can perform an asymptotic expansion around Minkowski space to perform this evolution, making the problem tractable. However, as discussed above, an expansion in the hyperboloidal coordinate $\rho$ is not appropriate. 

\begin{figure}
  \centering
  \begin{subfigure}[b]{0.49\textwidth}
    \centering
    \includegraphics{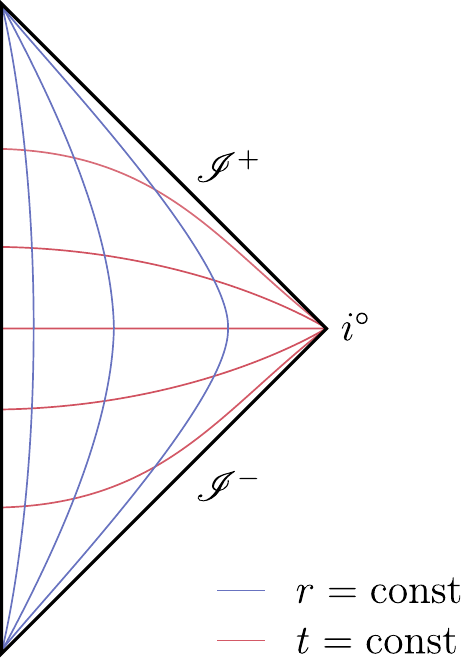}
    \caption{Spherical coordinate system}
  \end{subfigure}
  \begin{subfigure}[b]{0.49\textwidth}
    \centering
    \includegraphics{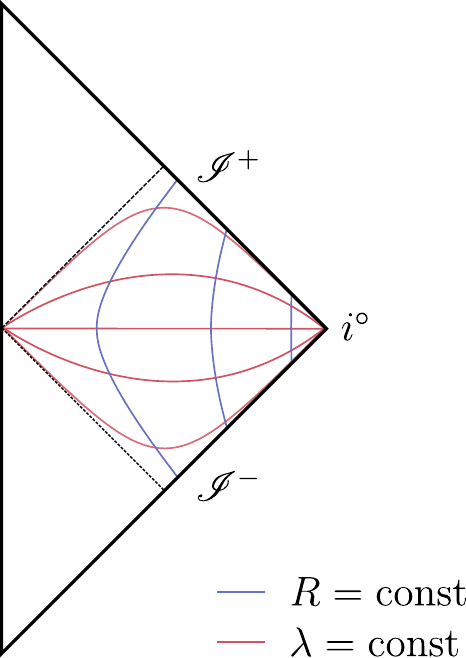}
    \caption{Friedrich coordinate system}
    \label{fig:Rlambda}
  \end{subfigure}
  \caption{Penrose diagram of Minkowski space with the standard $t,r$ coordinate systems in panel (a) and the coordinate system $\lambda, R$ used in this paper in panel (b). $\scrip$ and $\scrim$ are described by $\lambda=1$ and $\lambda=-1$ surfaces, and spatial infinity is described by the $R\to\infty$ limit. The dotted lines in panel (b) represent the boundary of the coordinate patch described by the Friedrich coordinates.}
  \label{fig:enter-label}
\end{figure}

Instead, we use Friedrich coordinates~\cite{FRIEDRICH199883, Friedrich:2002ru} for this analysis. Consider the coordinates $\lambda$ and $R$ defined on the $u<0$ and $v>0$ patch of Minkowski space as
\begin{subequations}
\begin{align}
  \label{eq:Rlambda_def}
  \lambda &= \frac{t}{r} \,, \\
  \frac{2}{R} &= \frac{1}{v} - \frac{1}{u} \,,
\end{align}
\end{subequations}
where $v=t+r$ and $u=t-r$. See Fig~\ref{fig:Rlambda} for an illustration of this coordinate system on a Penrose diagram for Minkowski space. First, note that at spatial infinity, we have $u\to-\infty$ and $v\to\infty$, implying that $R\to\infty$. Therefore, an expansion in large $R$ can approximate the metric near spatial infinity. In particular, at $t=0$ (equivalently $\lambda=0$) we have $R=r$. At $\scrip$, we have $v\to\infty$ and thus $R = -2u$. Thus, a large $R$ expansion corresponds to an early time expansion in $u$ as $u\to-\infty$. Similarly, at $\scrim$ we have $R=2v$, and a large $R$ expansion corresponds to an expansion in late advanced time $v$. 

Consider a metric $g^{(n)}_{ab}$ which is expanded around a Minkowski background and includes terms up to $1/R^{n}$ in a $1/R$ expansion. Let $g^{(n)}_{ab}$ satisfy the vacuum Einstein's equations up to this order, so that in a basis which is Cartesian at infinity the components of the Einstein tensor satisfy
\begin{equation}
  G_{ab}[g^{(n)}_{ab}] = O\left(\frac{1}{R^{n+3}}\right) \,,
\end{equation}
then $g^{(n)}_{ab}$ satisfies Einstein's equations approximately---to the given order---at spatial infinity \emph{as well as} at early times on $\scrip$, and at late times on $\scrim$.  As such, it provides an approximation of the metric in all three regimes, allowing us to evolve the metric all the way to null infinity in this expansion. 

Note however, that this expansion introduces a challenge. Suppose we truncate our $1/R$ expansion at to some order $N$ and manage to prove peeling at this order. At $\scrip$ this would mean that, for a metric expanded in $1/u$, terms up to $1/u^N$ are consistent with peeling. However, this is insufficient, as peeling should hold for the \emph{full} metric. Fortunately, we will see that only \emph{finitely many} terms in the $1/R$ (and thus $1/u$) expansion can violate peeling. This is crucial to our analysis, because this allows us to truncate at a finite order in $1/R$. Thus, we can make precise statements on peeling for the entire asymptotic expansion of the metric despite analyzing only a finite number of them.\footnote{Because our analysis is based on an asymptotic expansion, we cannot rule out non-perturbative terms with a vanishing asymptotic expansion at $\scrip$ that violate peeling. For space-times such as those in post-Minkowskian (PM) or post-Newtonian (PN) theory, it can be seen that such terms do not appear. More generally, a rigorous analysis of the PDEs is required to rule them out.}

This paper is structured as follows: In \S\ref{sec:scalar-field} we consider the simplest case of a massless scalar field on a Minkowski background, which has been studied in~\cite{Fuentealba:2024lll}. This simple example is a very useful precursor to the full GR equations, therefore for pedagogical reasons we include it in our paper as well. We solve this equation in a large $R$ expansion, and we study the limiting behaviour of these solutions at $\scrip$ and $\scrim$. In particular, we relate the initial data on a Cauchy slice to the asymptotics at null infinity, and identify precise symmetry criteria on the initial data for peeling to be satisfied. In \S\ref{sec:spin-s-field} we consider the massless spin-$s$ field on a Minkowski background. This is a useful step towards a study of the gravity case not only because the spin-2 field is identical to the linearized Weyl tensor on a Minkowski background, but also because we will see that some formulae for the general spin-$s$ case will be useful for future discussions of GR. We solve the field equations for the massless spin-$s$ field in the large $R$ expansion and study its asymptotics. We similarly relate the symmetries of the initial data to peeling for this field. In \S\ref{sec:gravity} we analyse the equations for gravity. At each order, we separate the solutions of the field equations into homogeneous and particular solutions. The homogeneous solutions are given by the spin-2 solutions given in \S\ref{sec:spin-s-field}. Using the properties of the Green's function, we study the asymptotics of the particular solutions and find symmetries of the initial data that guarantee peeling at null infinity at various orders. 

\section{Scalar field}
\label{sec:scalar-field}

In preparation for our discussion of the spin-$s$ wave equation, let us first consider the simplest case of the massless scalar field in some detail. Concretely, we will be considering solutions $\phi$ to the standard wave equation $\Box\phi = 0$ on flat space-time arising from an initial value problem with boundary conditions such that the scalar field $\phi$ vanishes asymptotically. The peeling property for a scalar field is that at $\scripm$, the scalar field $\phi$ should have the following behaviour
\begin{equation}
  \label{eq:phi-peel}
  \phi = \frac{\phi^\circ}{r} + o\left(\frac{1}{r}\right) \,.
\end{equation}
This will not generally be true for all solutions but constitutes a restriction on the class of initial data. In the following, we will relate the initial data at a $t=0$ time slice to the form of the scalar field at null infinity. In particular, this will allow us to map the condition \eqref{eq:phi-peel} to the initial data at $t=0$ slice. This will allow us to precisely state the criteria that the initial data must satisfy such that $\phi$ peels at $\scripm$ according to Eq.~$\eqref{eq:phi-peel}$.

Let us suppose that initial data on the $t = 0$ slice (or equivalently  the $\lambda = 0$ slice) is of the following form:
\begin{align}
  \label{eq:analytic_data}
  (\phi, \partial_\lambda\phi)|_{\lambda = 0} &= \Bigl(\sum_{n\geq0,\ell m}R^{-(n+1)}\mathbb{E}_{n\ell m}Y_{\ell m}, \sum_{n\geq0,\ell m}R^{-(n+1)}\mathbb{O}_{n\ell m}Y_{\ell m}\Bigr) \,.
\end{align}
Recall that at $t=0$, the coordinates $r$ and $R$ are identical. This is of course not the most general class of initial data that can be considered. Specifically, we are assuming that it can be analytically expanded in $1/R$. In fact, we expect that the conditions on the initial data can be significantly relaxed without affecting our result. However, because that complicates the discussion further, we will assume initial data of the form~\eqref{eq:analytic_data} here. This already consists of a very large class of physically interesting initial data. We will discuss some limitations of this data at the end of this section. Assuming that the initial data has a $1/r$ decay at spatial infinity of the form~\eqref{eq:analytic_data} does not trivially imply that $\phi$ has a $1/r$ decay at $\scripm$ as in~\eqref{eq:phi-peel}. In fact, we will see below that generally it doesn't.

We begin by solving the wave equation as an expansion in $1/R$. In $(\lambda, R)$-coordinates $\Box\phi = 0$ is given by:
\begin{align}
  \label{eq:scalar_wave}
  -\frac{(1 - \lambda^2)^2}{R^2}\Bigl[(1 - \lambda^2)\partial_\lambda^2 - 2\lambda R\partial_\lambda\partial_R - \frac{4}{1 - \lambda^2}R\partial_R - R^2\partial_R^2 - \Delta_{S^2}\Bigr]\phi &= 0 \,,
\end{align}
where $\Delta_{S^2}$ is the unit two-sphere Laplacian. The angular dependence of $\phi$ may be separated out by decomposing $\phi$ into spherical harmonics. Furthermore, because radial derivatives only appear in the combination $R\partial_R$, there exists solutions of the form $R^{-(1+n)}\phi_{n\ell m}(\lambda)Y_{\ell m}$. Plugging in this form of the solution into the wave equation \eqref{eq:scalar_wave} then gives us
\begin{align}
    \label{eq:jac_eq_scalar_wave}
    \Bigl[(1 - \lambda^2)\partial_\lambda^2 - 2(n + 1)\lambda\partial_\lambda + (\ell + n + 1)(\ell - n)\Bigr]\bigl[(1 - \lambda^2)^{-(n + 1)}\phi_{n\ell m}(\lambda)\bigr] &= 0 \,,
\end{align}
which may be recognized as the \textit{Gegenbauer equation}, or as a special case of the \textit{Hypergeometric} or \textit{Jacobi equation}. Alternatively, it can be shown to be equivalent to the \textit{associated Legendre equation}. In anticipation of generalizing the wave equation to higher spin in the next section, we will consider the more general Jacobi solution here. The Gegenbauer solutions to the scalar wave equation are studied in \cite{Fuentealba:2024lll}. Solutions $y_{\ell}^n(\lambda) := (1 - \lambda^2)^{-(n+1)}\phi_{n\ell m}(\lambda)$ are called \textit{Jacobi functions}. For an extensive list of formulas involving Jacobi functions see \S10 of \cite{bateman_bateman2}. For convenience, some relevant results are summarized in Appendix \ref{appendix:Jacobi}. For $\ell \geq n$, $y_\ell^n(\lambda)$ can be expressed as a linear combination of the associated Jacobi polynomial $P^{(n,n)}_{\ell-n}(\lambda)$ and the associated Jacobi function of the second kind $Q^{(n,n)}_{\ell-n}(\lambda)$. Jacobi functions have the following parity properties:
\begin{subequations}
\label{eq:Jacobi_parity}
  \begin{align}
    P_\nu^{(\alpha, \beta)}(-\lambda) &= (-1)^\nu P_\nu^{(\beta, \alpha)}(\lambda) \,, \\
    Q_\nu^{(\alpha, \beta)}(-\lambda) &= (-1)^{\nu+1}Q_\nu^{(\beta, \alpha)}(\lambda) \,.
  \end{align}
\end{subequations}
The coefficients of the even parity functions in the solution to the initial value problem \eqref{eq:analytic_data} must be proportional to $\mathbb{E}_{n\ell m}$, while the odd parity functions must be proportional to $\mathbb{O}_{n\ell m}$. Explicitly, a solution is given by:
\begin{subequations}
  \label{eq:scalar_Jacobi_sol}
  \begin{align}
    \phi_{n\ell m} &= (1 - \lambda^2)^{n+1}\bigl[\mathcal{A}_{n\ell m}P_{\ell-n}^{(n,n)}(\lambda) + \mathcal{B}_{n\ell m}Q_{\ell-n}^{(n,n)}(\lambda)\bigr] & &\textrm{for\;} \ell \geq n \,, \label{eq:scalar_Jacobi_sol_positive_nu} \\
    \phi_{n\ell m} &= (1 - \lambda^2)\bigl[\mathcal{A}_{n\ell m}(1 - \lambda)^nP_\ell^{(n,-n)}(\lambda) + \mathcal{B}_{n\ell m}(1 + \lambda)^nP_\ell^{(-n,n)}(\lambda)\bigr] & &\textrm{for\;} \ell < n \,, \label{eq:scalar_Jacobi_sol_negative_nu}
  \end{align}
\end{subequations}
where
\begin{subequations}
\label{eq:AB_even_positive_nu}
  \begin{align}
    \mathcal{A}_{n\ell m} &= \frac{\mathbb{E}_{n\ell m}}{P_{\ell-n}^{(n, n)}(0)} = \frac{(-1)^{\frac{\ell-n}{2}}}{2^n(\ell-n + 1)_n}\frac{(\ell+n)!!}{(\ell-n-1)!!}\mathbb{E}_{n\ell m} \,, \\
    \textrm{and} \quad \mathcal{B}_{n\ell m} &= \frac{\mathbb{O}_{n\ell m}}{(\partial_\lambda Q_{\ell-n}^{(n, n)})(0)} = (-1)^{\frac{\ell-n}{2}+1}8^n(\ell-n+1)_n\frac{\ell!^2(\ell-n-1)!!}{(\ell-n)!(\ell+n)!(\ell+n)!!}\mathbb{O}_{n\ell m} \,,
  \end{align}
\end{subequations}
for even $\ell-n \geq 0$, and
\begin{subequations}
\label{eq:AB_odd_positive_nu}
  \begin{align}
    \mathcal{A}_{n\ell m} &= \frac{\mathbb{O}_{n\ell m}}{(\partial_\lambda P_{\ell-n}^{(n, n)})(0)} = \frac{(-1)^{\frac{\ell-n-1}{2}}}{2^n(\ell-n)_{n+1}}\frac{(\ell+n-1)!!}{(\ell-n-2)!!}\mathbb{O}_{n\ell m} \,, \\
    \textrm{and} \quad \mathcal{B}_{n\ell m} &= \frac{\mathbb{E}_{n\ell m}}{Q_{\ell-n}^{(n, n)}(0)} = (-1)^{\frac{\nu-1}{2}}8^n(\nu)_{n+1}\frac{\ell!^2(\ell-n-2)!!}{(\ell-n)!(\ell+n)!(\ell+n-1)!!}\mathbb{E}_{n\ell m} \,,
  \end{align}
\end{subequations}
for odd $\ell-n \geq 0$. For $\ell < n$,
\begin{subequations}
\label{eq:AB_negative_nu}
  \begin{align}
    \mathcal{A}_{n\ell m} &= \frac{2^{n-1}\ell!\Gamma(\frac{1-\ell+n}{2})\Gamma(\frac{\ell+n}{2} + 1)}{\sqrt{\pi}(\ell+n)!}\mathbb{E}_{n\ell m} + \frac{2^{n}\ell!\Gamma(1 - \frac{\ell-n}{2})\Gamma(\frac{\ell+n+1}{2} + 1)}{\sqrt{\pi}(\ell+n+1)!}\mathbb{O}_{n\ell m} \,, \\
    \mathcal{B}_{n\ell m} &= \frac{(-1)^\ell2^{n-1}\ell!\Gamma(\frac{1-\ell+n}{2})\Gamma(\frac{\ell+n}{2} + 1)}{\sqrt{\pi}(\ell+n)!}\mathbb{E}_{n\ell m} - \frac{(-1)^\ell2^{n}\ell!\Gamma(1 - \frac{\ell-n}{2})\Gamma(\frac{\ell+n+1}{2} + 1)}{\sqrt{\pi}(\ell+n+1)!}\mathbb{O}_{n\ell m} \,.
  \end{align}
\end{subequations}

\subsection{Asymptotic behaviour}

Let us now study the asymptotic behaviour of the solution in the limits to $\scri$. Recall that $\scrip$ is described by the $\lambda\to1$ limit and $\scri^{-}$ is described by the $\lambda\to-1$ limit. Therefore, to obtain the limiting behaviour of the solutions to $\scrip$ and $\scrim$, we can analyse the asymptotic expansion of the solutions in this limit. Close to $\scri$ the solutions (\ref{eq:scalar_Jacobi_sol}) decay as 
\begin{subequations}
\begin{align}
  \phi_{n\ell m} &= (-1)^n\frac{(\ell+n)!}{\ell!n!}\mathcal{B}_{n\ell m}\Bigl(\frac{1 - \lambda^2}{R}\Bigr)^{n+1}\log\Bigl(\frac{1 - \lambda}{1 + \lambda}\Bigr) + O(1 - \lambda^2)^{n+1} & \textrm{for } \ell &\geq n \,, \\
  \phi_{n\ell m} &= \frac{(\ell+n)!}{\ell!n!}\frac{1-\lambda^2}{R^{n+1}}\Bigl[\mathcal{A}_{n\ell m}(1 - \lambda)^n + \mathcal{B}_{n\ell m}(1 + \lambda)^n\Bigr] + O(1 - \lambda^2)^2 & \textrm{for } \ell &< n \,. 
\end{align}
\end{subequations}
These expressions are easily converted to Bondi coordinates $u,v,r$ using
\begin{subequations}
\begin{align}
    1-\lambda &= \frac{|u|}{r} , \\
    1+\lambda &= \frac{v}{r} , \\
    \frac{1-\lambda^2}{R} &= \frac{1}{r} , 
\end{align}
\end{subequations}
so that a fall-off of $(1-\lambda)^p(1+\lambda)^q$ translates to a fall-off of $r^{-p}$ at $\scrip$ and $r^{-q}$ at $\scrim$. Explicitly,
\begin{subequations}
\begin{align}
  \phi_{n\ell m} &= (-1)^n\frac{(\ell+n)!}{\ell!n!}\frac{\mathcal{B}_{n\ell m}}{r^{n+1}}\log\left(\frac{|u|}{v}\right) + O\left(\frac{1}{r^{n+1}}\right) & \textrm{for } \ell &\geq n \,, \\
  \phi_{n\ell m} &= \frac{(\ell+n)!}{\ell!n!}\frac{\mathcal{A}_{n\ell m}}{rv^n} + \frac{(\ell+n)!}{\ell!n!}\frac{\mathcal{B}_{n\ell m}}{r|u|^n} + O\left(\frac{1}{r^2}\right) & \textrm{for } \ell &< n \,. 
\end{align}
\end{subequations}
In order for $\phi$ to peel, it needs to be of the order $1/r$, which will be the case if and only if $\mathbb{E}_{0\ell m} = 0$ for odd $\ell$ and $\mathbb{O}_{0\ell m} = 0$ for even $\ell$.

The condition for the $\log$ terms to vanish is equivalent to the $1/R$ part of $\phi$ having $\mathcal{PT}$-symmetry, where $\mathcal{PT}$ is a diffeomorphism that performs time reversal and the standard parity operator on the $2$-sphere given by
\begin{equation}
  \mathcal{PT}(\lambda, R, \theta, \phi)=(-\lambda, R, \pi-\theta,\pi+ \phi)\,.
\end{equation}
Thus, the condition on the initial data for peeling to occur is that the $1/R$ part of the scalar field $\phi$ is $\mathcal{PT}$-symmetric. We will see that this symmetry criterion shows up again for the spin-$s$ field and gravity. 

We now comment on the Poincar\'e invariance of the $\mathcal{PT}$-symmetry.  First, it is easy to see that Lorentz transformations that preserve the origin of our coordinate system leave the $\mathcal{PT}$ diffeomorphism invariant. Next, it can be seen that translations don't effect the leading order terms in $1/R$. Instead, they modify the subleading terms by a contribution proportional to the Lie derivative of the leading order term with respect to the translation. Therefore, the $\mathcal{PT}$ invariance of the leading order is Poincar\'e invariant.

It is also useful for future discussions to consider the Poincar\'e invariance of the $\mathcal{PT}$-symmetry of the subleading term of order $1/R^2$. Suppose that the leading term $\phi^\circ(\theta,\phi,\lambda)/R$ is 
$\mathcal{PT}$-symmetric. Then, under a translation $\xi^a$, the subleading term $\phi^{(1)}(\theta,\phi,\lambda)/R^2$ obtains a piece proportional to $\mathcal{L}_{\xi_\circ}\phi^\circ / R^2$, where $\xi_\circ^a = R \xi^a$ is the independent of $R$. Because $\xi^a$ is $\mathcal{PT}$-antisymmetric, this means that the subleading term $\phi^{(1)}(\theta,\phi,\lambda)$ obtains a $\mathcal{PT}$-antisymmetric piece under translations. Thus, if $\phi^{(1)}$ is $\mathcal{PT}$-antisymmetric, then this property is Poincar\'e invariant if $\phi^\circ$ is $\mathcal{PT}$-symmetric. 
\begin{remarks}\;
\begin{enumerate}
\item We only have to consider the leading order in the $1/R$ expansion because when the order in $1/R$ increases, the solutions~\eqref{eq:scalar_Jacobi_sol_positive_nu} also increase the powers of $1-\lambda^2$. Thus, they only contribute to $1/r^2$ and higher. We will later see similar results for the linear spin-$s$ fields, and for gravity, where only a finite number of terms in the expansion will contribute.
\item Notice that the solution \eqref{eq:scalar_Jacobi_sol} either peels or violates peeling at both $\scrip$ and $\scrim$ simultaneously. This is not a feature of the standard wave equation but rather a consequence of the initial data \eqref{eq:analytic_data}. This can be seen by considering spherically symmetric solutions to the standard wave equation \eqref{eq:scalar_wave}, which are given by $\phi = r^{-1}f(u) + r^{-1}g(v)$ in double null coordinates, where $f$ and $g$ are arbitrary. Choosing $f = 0$ and $g(v) = \log v$ results in a solution to the initial value problem
\begin{subequations}
  \begin{align}
    (\phi, \partial_\lambda\phi)|_{\lambda = 0} &= (R^{-1}\log(R), R^{-1}) \,, \\
    \textrm{or equivalently} \quad (\phi, \partial_t\phi)|_{t = 0} &= (r^{-1}\log(r), r^{-2}) \,,
  \end{align}
\end{subequations}
which peels at $\scrim$ but violates peeling at $\scrip$. This initial data is not of the form~\eqref{eq:analytic_data}. 
\end{enumerate}
\end{remarks}

\section{Asymptotic Expansion for spin-s massless fields}
\label{sec:spin-s-field}
In this section we will solve the general spin-$s$ massless free-field equation on flat space-time---which includes the scalar wave equation, the source free Maxwell equations, and the Linearized Einstein equations---and relate initial data to the peeling properties of the solution at $\scri$. A wonderfully efficient way to do this is to apply the Geroch-Held-Penrose (GHP) formalism \cite{Geroch:1973am, Penrose:1985bww, Penrose:1986ca}. 

The spin-$s$ massless field is described by $2s+1$ independent complex components $\phi_k$, where $0\leq k \leq 2s$. Each $\phi_k$ has both boost- and spin weights given by $s-k$. Therefore, $s$ is the spin weight of the $\phi_0$ component only, and should not be confused with the spin weight of the other $\phi_k$ components.

For example, the electromagnetic scalars are given by $\phi_0 = F_{ab}l^am^b$, $\phi_1 = \tfrac{1}{2}F_{ab}(l^a n^b +  \bar{m}^a m^b)$, and $\phi_2 = F_{ab}\bar{m}^a n^b$. In linearized gravity, the Weyl scalars satisfy the spin-2 massless wave equation where $\phi_k$ is given by $\Psi_k$, the NP Weyl scalar components.

The peeling criteria for a spin-$s$ field can be summarized as follows. We say that a spin-$s$ field $\phi_k$ peels at $\scrip$ if in the limit to $\scrip$ the components $\phi_k$ decay as
\begin{equation}
  \phi_k(u,r,\theta,\phi) = \frac{\phi_k^\circ(u,\theta,\phi)}{r^{2s+1-k}} + o\left( \frac{1}{r^{2s+1-k}}\right) \,,
\end{equation}
for large radius $r$. Similarly, we say that $\phi_k$ peels at $\scrim$ if in the limit to $\scrim$ the components $\phi_k$ decay as
\begin{equation}
  \phi_k(u,r,\theta,\phi) = \frac{\phi_k^\circ(u,\theta,\phi)}{r^{k+1}} + o\left( \frac{1}{r^{k+1}}\right) \,.
\end{equation}
The criteria can also be understood geometrically as a requirement on the smoothness of the spinor $\phi_{A_1\dots A_{2s}}$ at null infinity in the conformally completed space-time. Notice that in the $s=2$ case we get the same criteria as in Eq.~\eqref{eq:Psi_i_peel_scrip} and Eq.~\eqref{eq:Psi_i_peel_scrim}, and in the $s=0$ case we get the same criteria as in Eq.~\eqref{eq:phi-peel}.

We assume here the fields $\phi_k$ satisfies the massless free-field equations. In GHP form, on a \emph{generic} background, these are given by:
\begin{subequations}
  \label{eq:massless_GHP}
  \begin{align}
    (\textrm{\thorn} - [2s - k]\rho)\phi_{k+1} - (\eth' - [k + 1]\tau')\phi_k &= k\sigma'\phi_{k-1} - (2s - 1 - k)\kappa\phi_{k+2} \,, \\
    (\textrm{\thorn}' - [k + 1]\rho')\phi_k - (\eth - [2s - k]\tau)\phi_{k+1} &= (2s - 1 - k)\sigma\phi_{k+2} - k\kappa'\phi_{k-1} \,,
  \end{align}
\end{subequations}
where $0 \leq k \leq 2s - 1$, $\th$, $\th^\prime$, $\eth$ and $\eth^\prime$ are GHP derivative operators, and $\rho$, $\rho^\prime$, $\tau$, $\tau^\prime$, $\sigma$, $\sigma^\prime$, $\kappa$ and $\kappa^\prime$ are the GHP spin coefficients.

In this section we will work on a Minkowski background for simplicity. This simplifies Eq.~\eqref{eq:massless_GHP} significantly. Let us work in double null coordinates $u = \tfrac{1}{\sqrt{2}}(t-r), v = \tfrac{1}{\sqrt{2}}(t+r)$ such that the Minkowski metric is given by $\mathrm{d}s^2 = 2\mathrm{d}u\mathrm{d}v - r^2\mathrm{d}S^2$ where $\mathrm{d}S^2$ is the unit two-sphere metric. Let us also pick the natural tetrad $l^a\partial_a := \partial_v$ and $n^a\partial_a := \partial_u$, so that the only non-zero GHP spin-coefficients are $\rho$ and $\rho'$. Then, Eq.~\eqref{eq:massless_GHP} simplifies to:
\begin{subequations}
  \label{eq:massless_f}
  \begin{align}
    \textrm{\thorn}_c\phi_{k+1} - \eth'\phi_k &= 0 \,, \\
    \textrm{\thorn}'_c\phi_k - \eth\phi_{k+1} &= 0 \,, \label{eq:massless_b}
  \end{align}
\end{subequations}
where the operators $\textrm{\thorn}_c$ and $\textrm{\thorn}'_c$ are conformal GHP operators defined as
\begin{subequations}
\label{eq:conformal_commutators}
  \begin{align}
    \textrm{\thorn}_c\eta &= (\textrm{\thorn} - [s + 1 + \tfrac{1}{2}(p - q)]\rho)\eta \,, \\
    \textrm{\thorn}'_c\eta &= (\textrm{\thorn}' - [s + 1 - \tfrac{1}{2}(p - q)]\rho')\eta \,,
  \end{align}
\end{subequations}
where $p,q$ are the GHP weights of $\eta$ and $\frac{1}{2}(p - q)$ is the spin weight of $\eta$. The commutators for the operators $\textrm{\thorn}_c$, $\textrm{\thorn}'_c$, $\eth$ and $\eth'$  are given by
\begin{subequations}
  \label{eq:conformal_GHP_commutators}
  \begin{align}
    [\textrm{\thorn}_c, \textrm{\thorn}'_c] &= (p - q)\rho\rho' = [\eth, \eth'] \,, \\
    [\textrm{\thorn}_c, \eth] &= 0 \,.
  \end{align}
\end{subequations}
Using these commutators the massless free-field equations \eqref{eq:massless_f} can be decoupled by applying $\textrm{\thorn}_c$ to Eq.~\eqref{eq:massless_b}:
\begin{align}
  \label{eq:GHP_wave_a}
  0 = \th_c(\th'_c\phi_k - \eth\phi_{k+1}) &= \th_c\th'_c\phi_k - (\eth\th_c + [\th_c, \eth])\phi_{k+1} \nonumber \\
  &= \th_c\th'_c\phi_k - \eth\th_c\phi_{k+1} \nonumber \\
  &= (\th_c\th'_c - \eth\eth')\phi_k \,.
\end{align}
For convenience, we will define the spin-$s$ wave operator $\Ro := \th_c\th'_c - \eth\eth'$, generalizing the wave operator $\Box$. The angular dependence may be separated by expanding $\phi_k$ into spin weighted spherical harmonics:
\begin{align}
  \phi_k := \sum_{\ell m} \phi_{k,\ell m}\,\,{}_{s-k}Y_{\ell m} \,,
\end{align}
so that
\begin{align}
  \label{eq:massless_l_mode}
  \bigl[\textrm{\thorn}_c\textrm{\thorn}'_c - (\ell - s + k + 1)(\ell + s - k)\rho\rho'\bigr]\phi_{k,\ell m} &= 0 \,.
\end{align}

\subsection{The initial value problem}

Finally, let us generalize our solution to the initial value problem \eqref{eq:analytic_data} to massless free-fields of arbitrary spin. First, we will be considering the initial value problem $\Ro\phi_s = 0$, $(\phi_s, \partial_\lambda\phi_s)|_{\lambda = 0} = (\Phi, \Phi')$. The reason for choosing the $k = s$ component is that the solution to the massless free-field equation can be recovered from $\phi_s$ (cf. Eq.~\eqref{eq:ladder} below). Conversely, it is not possible, for instance, to obtain the $\ell = 0$ mode of $\phi_s$ from any of the other $\phi_{k\neq s}$ since these components have non-zero spin weight and therefore do not have any $\ell = 0$ modes themselves. We will once again consider data which can be expressed as a series in $1/R$:
\begin{align}
  \label{eq:spin_s_analytic_data}
  (\Phi, \Phi') &= \Bigl(\sum_{n,\ell\geq0}R^{-(s+n+1)}\mathbb{E}_{n\ell m}Y_{\ell m}, \sum_{n,\ell\geq0}R^{-(s+n+1)}\mathbb{O}_{n\ell m}Y_{\ell m}\Bigr) \,.
\end{align}
Writing $\phi_k = \sum_{n\geq0,\ell} R^{-(s+n+1)}\phi_{k,n\ell m}(\lambda)\,{}_{s-k}Y_{\ell m}$, $\Ro\phi_k=0$ becomes:
\begin{align}
\label{eq:spin_s_Jacobi}
  \Bigl[(1 - \lambda^2)\partial_\lambda^2 - 2((n + 1)\lambda - s + k)\partial_\lambda + (\ell + n + 1)(\ell - n)\Bigr]\bigl[(1 - \lambda^2)^{-(s + n + 1)}\phi_{k,n\ell m}(\lambda)\bigr] &= 0 \,.
\end{align}
Its solutions are given by
\begin{subequations}
  \label{eq:Jacobi_sol}
  \begin{align}
    \phi_{k,n\ell m} &= (1 - \lambda^2)^{s+n+1}\bigl[\mathcal{A}_{k,n\ell m}P_\nu^{(\alpha,\beta)}(\lambda) + \mathcal{B}_{k,n\ell m}Q_\nu^{(\alpha,\beta)}(\lambda)\bigr] & &\textrm{for\;} \ell \geq n \,, \label{eq:Jacobi_sol_positive_nu} \\
    \phi_{k,n\ell m} &= (1 - \lambda^2)^{s+n+1}\bigl[\mathcal{A}_{k,n\ell m}(1 + \lambda)^{-\beta}P_{\nu+\beta}^{(\alpha,-\beta)}(\lambda) + \mathcal{B}_{k,n\ell m}(1 - \lambda)^{-\alpha}P_{\nu+\alpha}^{(-\alpha,\beta)}(\lambda)\bigr] & &\textrm{for\;} \ell < n \,, \label{eq:Jacobi_sol_negative_nu}
  \end{align}
\end{subequations}
where the parameters $\alpha$, $\beta$, and $\nu$ are given by $k - s + n$, $s - k + n$, and $\ell - n$. As before, $P_{\nu}^{(\alpha,\beta)}(\lambda)$ is the Jacobi polynomial and  $Q_{\nu}^{(\alpha,\beta)}(\lambda)$ is the Jacobi function of the second kind. 
For $k = s$,
\begin{align}
    \Ro(2R^{-(s+n+1)}\phi_{s,n}) &= \Box(R^{-(n+1)}(1-\lambda^2)^{-s}\phi_{s,n}) \,,
\end{align}
(cf. Eq.~\eqref{eq:jac_eq_scalar_wave}) and the Jacobi solution \eqref{eq:Jacobi_sol} multiplied by $(1-\lambda^2)^{-s}$ reduces to the Jacobi solution \eqref{eq:scalar_Jacobi_sol} to the scalar wave equation. Hence, the coefficients $\mathcal{A}_{s,n\ell m}$ and $\mathcal{B}_{s,n\ell m}$ are given in terms of $\mathbb{E}_{n\ell m}$ and $\mathbb{O}_{n\ell m}$ by Eqs. \eqref{eq:AB_even_positive_nu}, \eqref{eq:AB_odd_positive_nu} and \eqref{eq:AB_negative_nu} of the previous section.

Suppose, now, that in addition to being solutions to the spin-$s$ wave equation \eqref{eq:GHP_wave_a}, the functions $\phi_k$ are also solutions to the massless free-field equations \eqref{eq:massless_f}. Given $\phi_s$, $\phi_k$ must satisfy
\begin{subequations}
\label{eq:ladder}
\begin{align}
    \phi_k &= (\eth'^\dagger\th_c)^{s-k}\phi_s \qquad \textrm{for } 0 \leq k \leq s \,, \\
    \phi_k &= (\eth^\dagger\th_c')^{k-s}\phi_s \qquad \textrm{for } s \leq k \leq 2s \,, 
\end{align}
\end{subequations}
where $\eth^\dagger$ and $\eth'^\dagger$ are the \textit{pseudo-inverse} operators of $\eth$ and $\eth'$, given by
\begin{subequations}
\label{eq:eth_dagger}
\begin{align}
    \eth^\dagger &: {}_sY_{\ell m} \longrightarrow -R(1-\lambda^2)^{-1}(\tfrac{1}{2}(\ell+s)(\ell-s+1))^{-\frac{1}{2}}{}_{s-1}Y_{\ell m} \,, \\
    \eth'^\dagger &: {}_sY_{\ell m} \longrightarrow R(1-\lambda^2)^{-1}(\tfrac{1}{2}(\ell-s)(\ell+s+1))^{-\frac{1}{2}}{}_{s+1}Y_{\ell m} \,.
\end{align}
\end{subequations}
Note that these are not inverse operators, since $\eth$ and $\eth'$ have kernels spanned by ${}_sY_{\ell=s}$ and ${}_sY_{\ell=-s}$ respectively. In addition to satisfying both Eq.~\eqref{eq:GHP_wave_a} and Eq.~\eqref{eq:ladder}, solutions have to satisfy the constraints
\begin{subequations}
\label{eq:constraints}
\begin{align}
    \th_c\phi_{k,\ell=s-k} &= 0 \,, & &\textrm{for } 1 \leq k \leq s \,, \\
    \th_c'\phi_{k,\ell=k-s} &= 0 \,, & &\textrm{for } s \leq k \leq 2s - 1 \,.
\end{align}
\end{subequations}
In $\lambda, R$ coordinates the operators $\th_c$ and $\th_c'$ are given by
\begin{subequations}
\begin{align}
    &R^{1+s+n}(1-\lambda^2)^{-(1+s+n)}\th_c(R^{-(1+s+n)}\phi_{k,n\ell m}) \nonumber \\
    &\qquad = R^{-1}(1-\lambda^2)\Bigl[(1-\lambda)\partial_\lambda - (k-s+n)\Bigr]\bigl[(1-\lambda^2)^{-(1+s+n)}\phi_{k,n\ell}\bigr] \,, \label{eq:thc} \\
    &\qquad = -(\ell-s+k)R^{-1}(1-\lambda^2)\bigl[\mathcal{A}_{k,n\ell m}P_\nu^{(\alpha-1,\beta+1)} + \mathcal{B}_{n\ell}Q_\nu^{(\alpha-1,\beta+1)}\bigr] \,, & &\textrm{if $\ell \geq n$} \,, \nonumber \\
    &R^{1+s+n}(1-\lambda^2)^{-(1+s+n)}\th_c'(R^{-(1+s+n)}\phi_{k,n\ell m}) \nonumber \\
    &\qquad = R^{-1}(1-\lambda^2)\Bigl[(1+\lambda)\partial_\lambda - (s-k+n)\Bigr]\bigl[(1-\lambda^2)^{-(1+s+n)}\phi_{k,n\ell m}\bigr] \,, \label{eq:thcp} \\
    &\qquad = -(\ell+s-k)R^{-1}(1-\lambda^2)\bigl[\mathcal{A}_{k,n\ell m}P_\nu^{(\alpha+1,\beta-1)} + \mathcal{B}_{k,n\ell m}Q_\nu^{(\alpha+1,\beta-1)}\bigr] \,, & &\textrm{if $\ell \geq n$} \,. \nonumber
\end{align}
\end{subequations}
For $\ell < n$, $\th_c(R^{-(1+s+n)}\phi_{k,n\ell m})$ and $\th_c'(R^{-(1+s+n)}\phi_{k,n\ell m})$ are more complicated, and most importantly, only vanish if $\phi_{k,n\ell m} = 0$. When $\ell \geq n$ the constraints \eqref{eq:constraints} are automatically satisfied. For the remaining coefficients, in order to satisfy the constraints \eqref{eq:constraints}, 
\begin{align}
    &\mathcal{A}_{k,n\ell m} = \mathcal{B}_{k,n\ell m} = 0 & &\textrm{for $\ell < n$ and $\ell < s$} \,.
\end{align}
Most of the remaining coefficients $\mathcal{A}_{k,n\ell m}$ and $\mathcal{B}_{k,n\ell m}$ can be computed from Eq.~\eqref{eq:ladder}. For example, for $\ell \geq n$,
\begin{align}
    &\mathcal{A}_{k,n\ell m}P_\nu^{(\alpha,\beta)} + \mathcal{B}_{k,n\ell m}Q_\nu^{(\alpha,\beta)} = (1-\lambda^2)^{-(1+s+n)}\phi_{k,n\ell m} \nonumber \\
    &\qquad = R^{1+s+n}(1-\lambda^2)^{-(1+s+n)}(\eth'^\dagger\th_c)^{s-k}(R^{-(1+s+n)}\phi_{s,n\ell m}) \\
    &\qquad = (-\sqrt{2})^{s-k}\sqrt{\frac{(\ell-s+k-1)_{s-k}}{(\ell+1)_{s-k}}}\bigl[\mathcal{A}_{s,n\ell m}P_\nu^{(\alpha,\beta)} + \mathcal{B}_{s,n\ell m}Q_\nu^{(\alpha,\beta)}\bigr] \,, \nonumber
\end{align}
so that
\begin{subequations}
  \begin{align}
    \mathcal{A}_{k,n\ell m} &= (-\sqrt{2})^{s-k}\sqrt{\frac{(\ell-s+k-1)_{s-k}}{(\ell+1)_{s-k}}}\mathcal{A}_{s,n\ell m} \,, \label{eq:Aknlm1} \\
    \mathcal{B}_{k,n\ell m} &= (-\sqrt{2})^{s-k}\sqrt{\frac{(\ell-s+k-1)_{s-k}}{(\ell+1)_{s-k}}}\mathcal{B}_{s,n\ell m} \,, 
  \end{align}
\end{subequations}
for $1 \leq k \leq s$, $\ell \geq n$ and similarly
\begin{subequations}
\begin{align}
    \mathcal{A}_{k,n\ell m} &= (\sqrt{2})^{k-s}\sqrt{\frac{(\ell+s-k-1)_{k-s}}{(\ell+1)_{s-k}}}\mathcal{A}_{s,n\ell m} \,, \label{eq:Aknlm2} \\
    \mathcal{B}_{k,n\ell m} &= (\sqrt{2})^{k-s}\sqrt{\frac{(\ell+s-k-1)_{s-k}}{(\ell+1)_{k-s}}}\mathcal{B}_{s,n\ell m} \,, 
\end{align}
\end{subequations}
for $s \leq k \leq 2s-1$, $\ell \geq n$. 

\subsection{Asymptotics}

\noindent Having obtained an explicit solution \eqref{eq:Jacobi_sol} to the initial value problem \eqref{eq:spin_s_analytic_data}, we may now easily find its asymptotics to study peeling using the expressions for the asymptotics of Jacobi functions found in appendix \ref{appendix:Jacobi}. In particular, similar to the scalar field, we obtain $\mathcal{PT}$-symmetry criteria for the solutions to peel at $\scrip$ and $\scrim$. Importantly, we find that only finitely many orders in $1/R$ can violate peeling. 

First, we translate the Friedrich coordinates of the solution~\eqref{eq:Jacobi_sol} to Bondi coordinates near $\scrim$. Close to $\scrim$, the solutions \eqref{eq:Jacobi_sol} decay as
\begin{subequations}
\label{eq:Jacobi_sol_asymptotics_scrim}
\begin{align}
\phi_{k,n\ell m} &\sim \mathcal{A}_{k,n\ell m}r^{-\max\{s+n+1,k+1\}} + \mathcal{B}_{k,n\ell m}r^{-\max\{s+n+1,k+1\}}\log r & &\textrm{for $\ell \geq n$} \,, \label{eq:Jacobi_sol_positive_nu_asymptotics_scrim} \\
\phi_{k,n\ell m} &\sim \mathcal{A}_{k,n\ell m}r^{-(k+1)} + \mathcal{B}_{k,n\ell m}r^{-(s+n+1)} & &\textrm{for $\ell < n$} \,, \label{eq:Jacobi_sol_negative_nu_asymptotics_scrim}
\end{align}
\end{subequations}
where $r$ is a radial Bondi coordinate. The fall-off of the solutions \eqref{eq:Jacobi_sol} near $\scrip$ is similarly given by
\begin{subequations}
\label{eq:Jacobi_sol_asymptotics_scrip}
\begin{align}
\phi_{k,n\ell m} &\sim \mathcal{A}_{k,n\ell m}r^{-\max\{s+n+1,2s+1-k\}} + \mathcal{B}_{k,n\ell m}r^{-\max\{s+n+1,2s+1-k\}}\log r & &\textrm{for $\ell \geq n$} \,,  \label{eq:Jacobi_sol_positive_nu_asymptotics_scrip} \\
\phi_{k,n\ell m} &\sim \mathcal{A}_{k,n\ell m}r^{-(s+n+1)} + \mathcal{B}_{k,n\ell m}r^{-(2s+1-k)} & &\textrm{for $\ell < n$} \,. \label{eq:Jacobi_sol_negative_nu_asymptotics_scrip}
\end{align}
\end{subequations}
The asymptotic behaviour of the solution can now be read off from Eq.~\eqref{eq:Jacobi_sol_asymptotics_scrim} and Eq.~\eqref{eq:Jacobi_sol_asymptotics_scrip}. Recall that a massless field $\phi_k$ peels at $\scrim$ if $\phi_k = O(1/r^{k+1})$. The $\mathcal{A}_{k,n\ell m}$ terms in both Eq.~\eqref{eq:Jacobi_sol_positive_nu_asymptotics_scrim} and Eq.~\eqref{eq:Jacobi_sol_negative_nu_asymptotics_scrim} are easily seen to peel. Additionally, since if $\ell < n$ it follows that $s + n + 1 > s + \ell + 1 \geq s + |s-k| + 1 \geq k + 1$, so that the $\mathcal{B}_{k,n\ell m}$ term in~\eqref{eq:Jacobi_sol_negative_nu_asymptotics_scrim} also peels. If $\ell \geq n$, the $\mathcal{B}_{k,n\ell m}$ solutions in~\eqref{eq:Jacobi_sol_positive_nu_asymptotics_scrim} peel if $s + n + 1 > k + 1$, which will be the case if $n > s$. Hence, the only term that violate peeling is the $\mathcal{B}_{k,n\ell m}$ term in~\eqref{eq:Jacobi_sol_positive_nu_asymptotics_scrim}, and only for the finitely many orders where $n\leq s$. 

Thus, the  solutions \eqref{eq:Jacobi_sol} peel if $\mathbb{E}_{n\ell m} = 0$ for odd $\ell - n \geq 0$ and $\mathbb{O}_{n\ell m} = 0$ for even $\ell - n \geq 0$ where $n\leq s$. Under a $\mathcal{PT}$-transformation, $\phi_k$ transforms for integer $s$ as
\begin{align}
    \label{eq:pullback_NP_scalar}
    \mathcal{PT}[\phi_k](\lambda,R,\theta,\phi) &= (-1)^s\phi_{2s-k}(-\lambda, R, \pi - \theta, \pi + \phi) \,.
\end{align}
If $\mathcal{PT}[\phi_k](\lambda,R,\theta,\phi) = (-1)^s\phi_{2s-k}(\lambda,R,\theta,\phi)$ we say $\phi_k$ is $\mathcal{PT}$-symmetric and has $\mathcal{PT}$-parity $1$ and if $\mathcal{PT}[\phi_k](\lambda,R,\theta,\phi) = -(-1)^s\phi_{2s-k}(\lambda,R,\theta,\phi)$ we say $\phi_k$ is $\mathcal{PT}$-antisymmetric and has $\mathcal{PT}$-parity $-1$. Hence, we can state an alternative peeling condition as follows: \emph{$\phi_k$ peels if and only if at order $1/R^{1+s+n}$, the $\ell \geq n$ modes of $\phi_k$ have a $(-1)^{s+n}$ parity under the $\mathcal{PT}$ operator, for $n \leq s$}.\footnote{Following the discussion for the scalar field, this property can be verified to be invariant under Poincar\'e transformations}

Without the $\mathcal{PT}$ conditions, the solutions fall-off as
\begin{subequations}
\begin{align}
\label{eq:peeling_no_sym}
\phi_k &= O\left(\frac{1}{r^{k+1}}\right) & &\textrm{for $k < s$} \,, \\
\phi_k &= O\left(\frac{\log r}{r^{k+1}}\right) & &\textrm{for $k \geq s$} \,.
\end{align}
\end{subequations}
The results for $\scrip$ follow analogously to the results for $\scrim$. Similar to the scalar field case, solutions \eqref{eq:Jacobi_sol} either peel or violate peeling at both $\scrip$ and $\scrim$. 

\begin{remarks}\;
\begin{enumerate}
\item Similar to the scalar field case, there exist solutions that peel at either $\scrip$ or $\scrim$. For example, the solutions (see Appendix~\ref{appendix:gen_sol} for a derivation)
\begin{align}
    \phi_{k,\ell m} &= r^{-1-s-\ell}(r^2\partial_v)^{\ell+s-k}V_{\ell m}(v) + r^{-1-s-\ell}(r^2\partial_u)^{\ell-s+k}U_{\ell m}(u) \,,
\end{align}
where $V_{\ell m}(v)$ and $U_{\ell m}(u)$ are arbitrary, may fail to peel. Here $r=(v-u)/2$. In the simple case of $k = 0$, $\ell = s$, $V_{\ell m} = 0$, and $U_{\ell m} = |u|^n\delta_{\ell,s}\delta_{m,0}$ we obtain the solution
\begin{align}
    \phi_0 &= r^{-1-2s}|u|^n{}_sY_{s0} \,,
\end{align}
which peels at $\scrim$, but which does not peel at $\scrip$ if $n > 2s$. In $\lambda,R$ coordinates this solution is given by
\begin{align}
    \phi_0 &= R^{-1-2s+n}(1-\lambda)^{1+2s-n}(1+\lambda)^{1+2s}{}_sY_{s0} \,.
\end{align}
For $n > s$, these solutions do not arise from the initial data \eqref{eq:spin_s_analytic_data} considered in this section. 
\item Instead of imposing the $\mathcal{PT}$-parity condition at all orders $n\leq s$, we can impose it for $n\leq w < s$ for some $w$. In this case not all $\phi_k$ peel, but the situation is improved compared to~\eqref{eq:peeling_no_sym}. In particular, we get that along $\scrim$ we have
\begin{subequations}
\begin{align}
\phi_k &= O\left(\frac{1}{r^{k+1}}\right) \quad \textrm{for $k \leq s+w$} \,, \\
\phi_k &= O\left(\frac{\log r}{r^{k+1}}\right) \quad \textrm{for $k > s+w$} \,.
\end{align}
\end{subequations}
\item Under a $\mathcal{PT}$ transformation the tetrad vectors transform as
\begin{subequations}
\begin{align}
    \mathcal{PT}_*[l^a] &= \mathcal{PT}_*[\partial_v^a] = -\partial_u^a = -n^a \,, \\
    \mathcal{PT}_*[m^a] &= \mathcal{PT}_*\Bigl[\frac{1}{\sqrt{2}r}\bigl(\partial_\theta^a + \frac{i}{\sin\theta}\partial_\phi^a\bigr)\Bigr] = \frac{1}{\sqrt{2}r}\bigl(-\partial_\theta^a + \frac{i}{\sin(\pi - \theta)}\partial_\phi^a\bigl) = -\bar{m}^a \,,
\end{align}
\end{subequations}
and since $\mathcal{PT} = \mathcal{PT}^{-1}$, $\mathcal{PT}_*[n^a] = -l^a$ and $\mathcal{PT}_*[\bar{m}^a] = -m^a$. From this we can compute, for example,
\begin{align}
\begin{split}
    \mathcal{PT}[\Psi_1](\lambda, R, \theta, \phi) &= \mathcal{PT}[C_{abcd}l^am^bl^cn^d](\lambda, R, \theta, \phi) \\
    &= C_{abcd}(-n^a)(-\bar{m}^b)(-n^c)(-l^d)(-\lambda, R, \pi - \theta, \pi + \phi) \\
    &= \Psi_3(-\lambda, R, \pi - \theta, \pi + \phi) \,.
\end{split}
\end{align}
Similarly, it can be shown that for a general spin-$s$ field, $\phi_k$ transforms according to Eq.~\eqref{eq:pullback_NP_scalar}. 
\end{enumerate}
\end{remarks}

\section{Asymptotic Expansion for gravity}
\label{sec:gravity}
We have found explicit solutions to the spin-$s$ massless free-field equations \eqref{eq:massless_f}. For spin-2 massless fields, the massless free-field equations may be identified with the linearized Bianchi identities, so that the functions $\{\phi_k\}$ correspond to components of the linearized Weyl tensor. In this section, we investigate the asymptotics of the full Bianchi identities by promoting the linear wave equations $\Ro\phi_k = 0$ to non-linear wave equations for the Weyl scalars $\{\Psi_k\}$, which hold in curved space-times. Of course, there is little hope of finding an explicit solution to such a problem. Instead, we make an additional assumption on the regularity of the metric which lets us expand the wave equation in powers of $R^{-1}$ near $\inot$. Our wave equation becomes a series of inhomogeneous wave equations of the form $\Ro(R^{-3-n}\Psi^{(n)}_k) = R^{-5-n}S_k^{(n)}$, where the source $S_k^{(n)}$ depends on the lower order solutions $\Psi^{(<n)}_k$. We are then able to show that assuming Peeling holds at the linear level, then all Weyl scalars except possibly $\Psi_0$ can be made to peel at $\scrip$ with a good choice of initial data. 

\subsection{Asymptotically regular space-times}
\label{sec:Asympt_regular_spacetimes}
Similar to the discussion above for the scalar and spin-$s$ fields, the asymptotic analysis for gravity will require an assumption on the asymptotic form of the metric. In this section we define the assumptions on this form of the metric that we make in this paper. First, let us introduce some notation for asymptotic expansion of tensors. 

Let us start by recalling the definition of $O_\infty$: A function $f(R)$ is $O_\infty(1/R^n)$ if the function $f(R)$ is $O(1/R^n)$ and $\partial_R^m f(R)$ is $O(1/R^{m+n})$ for all integers $m$. To extend this to tensors, we need to pick a preferred coordinate system and apply the definition for scalars on the \emph{components} of the tensor. Because we will be expanding around a Minkowski background $\eta_{ab}$, we can use this metric to pick such a coordinate system. We say that a tensor $t_{a_1\ldots a_n}$ is $O_\infty(1/R^n)$ if all of its components are $O_\infty(1/R^n)$ in a Cartesian coordinate system of the background Minkowski metric $\eta_{ab}$.

We now describe the asymptotic assumptions we make on the metric. We assume that there exists a background Minkowski metric $\eta_{ab}$ such that the metric $g_{ab}$ admits an asymptotics expansion around $\eta_{ab}$ of the form
\begin{equation}
    \label{eq:asymptotic-regular-def}
     g_{ab} = \eta_{ab} + \frac{h_{ab}}{R} + \sum_{n\geq2} \frac{h_{ab}^{(n)}}{R^n} \,,
\end{equation}
where the coordinate $R$ is defined using Eq.~\eqref{eq:Rlambda_def} with respect to some rest frame of the background metric $\eta_{ab}$. Note that the existence of an expansion~\eqref{eq:asymptotic-regular-def} is reference frame independent. Written in terms of initial data on a Cauchy slice, we assume that the three metric $\gamma_{ab}$ and the extrinsic curvature $K_{ab}$ are of the form
\begin{align}
    \gamma_{ab} &= \delta_{ab} + \sum_{n\geq1} \frac{\gamma_{ab}^{(n)}}{R^n} , \\
    K_{ab} &= \sum_{n\geq1} \frac{K_{ab}^{(n)}}{R^{n+1}} \,.
\end{align}
 We will additionally make the assumption (in addition to~\eqref{eq:asymptotic-regular-def}) that
\begin{align}
    g_{ab} - \eta_{ab} = O(1-\lambda^2) \label{eq:metric_regularity_scri}\,.
\end{align}
This condition ensures that the background Minkowski metric $\eta_{ab}$ chosen to expand around is well-behaved at null infinity, and that null infinity has a well-defined gravitational radiation field. This condition implies that $\Psi_4$ will peel at $\scrip$, and similarly that $\Psi_0$ will peel at $\scrim$.  

A metric $g_{ab}$ which admits a background metric $\eta_{ab}$ such that $g_{ab}$ admits an expansion of the form~\eqref{eq:asymptotic-regular-def} and satisfies~\eqref{eq:metric_regularity_scri} will be said to be asymptotically regular.

\subsection{Wave equation for the Weyl Scalars}
From the asymptotic form of the metric \eqref{eq:asymptotic-regular-def} it follows that $C_{abcd} = O_\infty(R^{-3})$, and that $\nabla_{a}C_{bcde} = \mathring{\nabla}_aC_{bcde} + O_\infty(R^{-5}) = O_\infty(R^{-4})$,
where $\mathring{\nabla}$ denotes the flat covariant derivative. Expanding the Bianchi identities around the flat metric yields
\begin{align}
    \label{eq:Penrose_wave}
    0 = \nabla_{[a}C_{bc]de} &= \mathring{\nabla}_{[a}C_{bc]de} + (\nabla-\mathring{\nabla})_{[a}C_{bc]de} = \mathring{\nabla}_{[a}C_{bc]de} + O_\infty(R^{-5}) \,.
\end{align}
Hence, the Weyl scalars satisfy 
\begin{align}
\Roo\Psi_k &= O_\infty(R^{-6}) \,.
\end{align}
At this point, after expanding $\Psi_k = \sum^{\infty}_{n\geq0}\Psi_{k,n}R^{-3-n}$,  we obtain at each order a flat spin-2 wave equation, identical to the one discussed in the previous section, with a source that depends on the lower order solutions. The two lowest order equations are given by:
\begin{subequations}
\begin{align}
\Roo(R^{-3}\Psi_{k,0}) &= 0 \,, \label{eq:leading_order_EFE} \\
\Roo(R^{-4}\Psi_{k,1}) &= R^{-6}S_{k,1}(h_{ab}) \,.
\end{align}
\end{subequations}
The homogeneous solution is given by the spin-2 solution \eqref{eq:Jacobi_sol}. As a quick summary, the conditions for the components $\{\psi_k\}$ of a spin-2 field satisfying $\Roo\psi_k = 0$ to peel at $\scrip$ are as follows: 
\begin{subequations}
\begin{align}
\psi_4 &= \psi_4^\circ r^{-1} + o(r^{-1}) & &\textrm{always} \,, \\
\psi_3 &= \psi_3^\circ r^{-2} + o(r^{-2}) & &\textrm{always} \,, \\
\psi_2 &= \psi_2^\circ r^{-3} + o(r^{-3}) & &\textrm{if} \quad \mathcal{B}_{k,0\ell m} = 0 \,, \\
\psi_1 &= \psi_1^\circ r^{-4} + o(r^{-4}) & &\textrm{if} \quad \mathcal{B}_{k,0\ell m} = \mathcal{B}_{k,1\ell m} = 0 \, , \\
\psi_0 &= \psi_0^\circ r^{-5} + o(r^{-5}) & &\textrm{if} \quad \mathcal{B}_{k,0\ell m} = \mathcal{B}_{k,1\ell m} = \mathcal{B}_{k,2\ell m} = 0 \,.
\end{align}
\end{subequations}
The asymptotic behaviour at $\scrim$ is similar, but with $\psi_{4-k}$ replacing $\psi_k$. Notice that only a finite number of terms in the $1/R$ expansion can break peeling! We also recall that the vanishing of the $\mathcal{B}_{k,n\ell m}$ terms is intimately related to the $\mathcal{PT}$-symmetry of the solutions. Specifically, $\mathcal{B}_{k,n\ell m}$ vanishes if the $n$-th order term has a $\mathcal{PT}$ parity of $(-1)^n$.

Explicitly, with the assumption $\mathcal{B}_{k,0\ell m} = \mathcal{B}_{k,1\ell m} = 0$, the asymptotic Weyl scalars are given in Bondi coordinates by
\begin{subequations}
\label{eq:leading_psi:asymptotic_form}
\begin{align}
    \psi_4^\circ &= \sum_{n\geq 3, n-1\geq\ell}\frac{(-1-n)_{\ell+2}}{(\ell+2)!}\mathcal{B}_{4,n\ell m}|u|^{-2-n}{}_{-2}Y_{\ell m} \,, \\
    \psi_3^\circ &= \sum_{n\geq 2,n-1\geq\ell\geq2}\frac{(-n)_{\ell+1}}{(\ell+1)!}\mathcal{B}_{3,n\ell m}|u|^{-1-n}{}_{-1}Y_{\ell m} \,, \\
    \psi_2^\circ &= \sum_{\ell\geq0}\mathcal{A}_{2,0\ell m}\,Y_{\ell m} + \sum_{n\geq 1,n-1\geq\ell\geq2}\frac{(1-n)_\ell}{\ell!}\mathcal{B}_{2,n\ell m}|u|^{-n}Y_{\ell m} \,, \\
    \psi_1^\circ &= \mathcal{A}_{1,11m}\;{}_1Y_{1m} + \sum_{\ell\geq 2}(\tfrac{1}{2}(\ell+2)\mathcal{A}_{1,0\ell m}u + \mathcal{A}_{1,1\ell m})\,{}_1Y_{\ell m} \nonumber \\
    &\phantom{= \mathcal{A}_{1,11m}\;{}_1Y_{1m}\,\,} + \sum_{n\geq 2, n-1\geq\ell\geq2}\frac{(2-n)_{\ell-1}}{(\ell-1)!}\mathcal{B}_{2,n\ell m}|u|^{1-n}{}_1Y_{\ell m} \,.
\end{align}
\end{subequations}

\subsection{The particular solution}
The full solution to Eq.~\eqref{eq:Penrose_wave} is given by
\begin{align}
    \Psi_{k,n\ell m} &= \psi_{k,n\ell m} + \Upsilon_{k,n\ell m} \,,
\end{align}
where $\psi_{k,n\ell m}$ is homogeneous solution, and where a particular solution $\Upsilon_{k,n\ell m}$ for $\nu \geq 0$ is given by
\begin{align}
    \label{eq:particular_solution}
    \Upsilon_{k,n\ell m} &= \frac{(-1)^\alpha\nu!(\nu+\alpha + \beta)!}{2^{\alpha+\beta}(\nu+\alpha)!(\nu+\beta)!}(1 - \lambda^2)^{n+3}\Bigl[P_\nu^{(\alpha,\beta)}(\lambda)\mathcal{Q}_{k,n\ell m}(\lambda) + Q_\nu^{(\alpha,\beta)}(\lambda)\mathcal{P}_{k,n\ell m}(\lambda)\Bigr] \,,
\end{align}
and for $\nu < 0$ by
\begin{align}
    \label{eq:particular_solution2}
    \Upsilon_{k,n\ell m} &= C(1 - \lambda^2)^{n+3}\Bigl[(1-\lambda)^{-\alpha}P_{\nu+\alpha}^{(-\alpha,\beta)}(\lambda)\mathcal{I}_{k,n\ell m}(\lambda) + (1+\lambda)^{-\beta}P_{\nu+\beta}^{(\alpha,-\beta)}(\lambda)\mathcal{J}_{k,n\ell m}(\lambda)\Bigr] \,,
\end{align}

\noindent where $C$ is a numerical constant and $\mathcal{P}_{k,n\ell m}$, $\mathcal{Q}_{k,n\ell m}$, $\mathcal{I}_{k,n\ell m}$, and $\mathcal{J}_{k,n\ell m}$ are given by
\begin{subequations}
\label{eq:Greens_integrals}
\begin{align}
    \mathcal{P}_{k,n\ell m} &= \int_{-1}^\lambda P_\nu^{(\alpha,\beta)}(\lambda')S_{k,n\ell m}(\lambda')(1 - \lambda')^{k-4}(1 + \lambda')^{-k}\mathrm{d}\lambda' \,, \label{eq:P_Greens_integral} \\
    \mathcal{Q}_{k,n\ell m} &= \int_0^\lambda Q_\nu^{(\alpha,\beta)}(\lambda')S_{k,n\ell m}(\lambda')(1 - \lambda')^{k-4}(1 + \lambda')^{-k}\mathrm{d}\lambda' \,, \\
    \mathcal{I}_{k,n\ell m} &= \int_0^\lambda P_{\nu+\beta}^{(\alpha,-\beta)}(\lambda')S_{k,n\ell m}(\lambda')(1 - \lambda')^{k-4}(1 + \lambda')^{-2-n}\mathrm{d}\lambda' \,, \\
    \mathcal{J}_{k,n\ell m} &= \int_0^\lambda P_{\nu+\alpha}^{(-\alpha,\beta)}(\lambda')S_{k,n\ell m}(\lambda')(1 - \lambda')^{-2-n}(1 + \lambda')^{-k}\mathrm{d}\lambda' \,. \label{eq:J_integral}
\end{align}
\end{subequations}
For the moment, we shall presume that the lower limit $-1$ on the first integral exists, and return to the question whether this is true or not later. There are two ways in which the particular solution can fail to peel. In the first place, we need to make sure that $\mathcal{Q}_{k,n\ell m}$ does not diverge as $\lambda \to 1$. Secondly, even if $\mathcal{P}_{k,n\ell m}$ does not diverge as $\lambda \to 1$, the logarithm in $Q^{(\alpha,\beta)}_\nu$ may still break the peeling of $\Psi_0$ and $\Psi_1$ at $\scrip$ if the integral is non-zero. In order for the solution to peel, we need $\mathcal{P}_{k,n\ell m}$ to have a vanishing limit of order $O(1/\log(1-\lambda))$ for $n = 1$. Finally, we need $\mathcal{I}_{k,n\ell m}$ to not diverge and $\mathcal{J}_{k,n\ell m}$ to blow up no faster than $(1-\lambda)^{2-k-n}$. Summarizing, showing that
\begin{subequations}
\label{eq:integral_decay}
\begin{align}
    \mathcal{Q}_{k,n\ell m} &= O(1) \,, \label{eq:Q_integral_decay} \\
    \mathcal{P}_{k,n\ell m} &= O(1) \,, \label{eq:P_integral_decay2} \\
    \mathcal{P}_{k,1\ell m} &= O(1/\log(1-\lambda)) \,, \label{eq:P_integral_decay} \\
    \mathcal{I}_{k,n\ell m} &= O(1) \,, \label{eq:I_integral_decay} \\
    \mathcal{J}_{k,n\ell m} &= O\bigl((1-\lambda)^{2-k-n}\bigr) \,, \label{eq:J_integral_decay}
\end{align}
\end{subequations}
would guarantee peeling for $\Psi_1$. The conditions \eqref{eq:integral_decay} fall into one of the following three categories: 
\begin{enumerate}
    \item \label{enum:peeling_1} The $P_{\nu\geq0}^{(\alpha,\beta)}$, $Q_{\nu\geq2}^{(\alpha,\beta)}$, $(1-\lambda)^{-\alpha}P_{\nu+\alpha<\alpha}^{(-\alpha,\beta)}$ and $(1+\lambda)^{-\beta}P_{\nu+\beta<\beta}^{(\alpha,-\beta)}$ solutions to the Jacobi equation all peel for $k\neq0$. The conditions \eqref{eq:Q_integral_decay}, \eqref{eq:P_integral_decay2}, and \eqref{eq:I_integral_decay} state that if their coefficients are finite, then the corresponding components of the particular solution have the same (peeling) fall-off rate. 
    \item \label{enum:peeling_2} 
    The solution $(1+\lambda)^{-\beta}P^{(\alpha,-\beta)}_{\nu+\beta<\beta}$ corresponds to a Weyl scalar having the fall-off rate $(1-\lambda)^{n+3}$, so that the integral $\mathcal{J}_{k,n\ell m}$ \eqref{eq:J_integral} must blow up no faster than $(1-\lambda)^{2-k-n}$. 
    \item \label{enum:peeling_3} The solution $Q^{(\alpha,\beta)}_{\nu=1}$ corresponds to a Weyl scalar $\Psi_1$ that falls off as $(1-\lambda)^{4}\log(1-\lambda)$, so that the corresponding particular solution only peels if its coefficient $\mathcal{P}_{k,1\ell m}$ \eqref{eq:P_Greens_integral} decays at least as fast as $1/\log(1-\lambda)$. 
\end{enumerate}
The conditions in category \ref{enum:peeling_1} and \ref{enum:peeling_2} are conditions on every coefficient in the $1/R$ expansion. \\
\\
In order for $\Psi_1$ to peel it would be sufficient to show that
\begin{align}
    \mathcal{P}_{2,1\ell m} = O_1(1-\lambda) \,,
\end{align}
instead of \eqref{eq:P_integral_decay2}, since $\Psi_1$ can be recovered from $\Psi_2$ from the Bianchi identities. The reason for the stronger decay requirement is that obtaining $\Psi_1$ from $\Psi_2$ involves taking a derivative of the form $(1-\lambda)\partial_\lambda$, so Eq.~\eqref{eq:P_integral_decay} for $k=2$ is not enough. The $\lambda$ derivatives of the log terms will reduce the $1-\lambda$ weight, and break peeling for $\Psi_1$. The behaviour of these integrals is governed by the source term $S_{k,n\ell m}$. This source term involves the metric, so that computing $S_{k,n\ell m}$ explicitly would involve solving another set of second order differential equations coupled to $\Psi_k$, or reconstructing the metric from the $\Psi_k$. Fortunately, there is another way that avoids these complications: we will show that there exists a gauge in which all $\Psi_k$, except $\Psi_0$, peel at $\scrip$. To start, let us consider the leading order solution $\Psi_{k,0\ell m}$ that is $\mathcal{PT}$-symmetric, such that $\mathcal{B}_{k,0\ell m} = 0$, which is simply given by (cf. Eq.~\eqref{eq:Jacobi_sol}, Eq.~\eqref{eq:leading_order_EFE})
\begin{align}
\label{eq:leading_order_peeling}
\Psi_{k,0\ell m} &= \mathcal{A}_{k,0\ell m}(1 - \lambda^2)^3P_\ell^{(k-2,2-k)}(\lambda) \,,
\end{align}
If $S_{k,1}$ were $\mathcal{PT}$-symmetric, then
\begin{align}
    \mathcal{P}_{2,1\ell m}(1) &= \int_{-1}^1P_{\ell-1}^{(1,1)}(\lambda')S_{2,1\ell m}(\lambda')(1 - \lambda'^2)^{-2}\mathrm{d}\lambda' \nonumber \\
    &= -\int_1^{-1}P_{\ell-1}^{(1,1)}(-\lambda')S_{2,1\ell m}(-\lambda')(1 - \lambda'^2)^{-2}\mathrm{d}\lambda' \\
    &= \int_{-1}^1(-1)^{\ell-1}P_{\ell-1}^{(1,1)}(\lambda')(-1)^\ell S_{2,1\ell m}(\lambda')(1 - \lambda'^2)^{-2}\mathrm{d}\lambda' = -\mathcal{P}_{2,1\ell m}(1) \,. \nonumber
\end{align}
Hence $\mathcal{P}_{2,1\ell m}(1) = 0$. Let us see what the $\mathcal{PT}$-symmetry of $\Psi_{k,0}$ implies for the metric. Since $\Psi_{k,0}$ is a linear function of $h_{ab}$, it follows that
\begin{align}
    \Psi_{k,0}(h_{ab}) &= \mathcal{PT}[\Psi_{4-k,0}(h_{ab})] = \Psi_{k,0}(\mathcal{PT}[h_{ab}]) \,,
\end{align}
and hence
\begin{align}
    \mathcal{PT}[h_{ab}] - h_{ab} = \Delta h_{ab} \,,
\end{align}
where $\Psi_{k,0}(\Delta h_{ab}) = 0$, so that the metric $\eta_{ab} + \Delta h_{ab}R^{-1} + O(R^{-2})$ has vanishing curvature to subleading order. It can then be shown that $\Delta h_{ab}$ can be made to vanish through a gauge transformation (see the $M \to 0$ limit of~\cite{Price:2006ke}). In a gauge where $h_{ab}$ is $\mathcal{PT}$-symmetric, $S_{k,1}$ will also be $\mathcal{PT}$-symmetric. This follows from the fact that $S_{k,1}$ is a linear function of $h_{ab}$, so that
\begin{align}
    S_{k,1}(h_{ab}) &= S_{k,1}(\mathcal{PT}[h_{ab}] - \Delta h_{ab}) = S_{k,1}(\mathcal{PT}[h_{ab}]) = \mathcal{PT}[S_{k,1}(h_{ab})] \,.
\end{align}
It remains to show that $\mathcal{P}_{2,n\ell m}$ is differentiable at $\scrip$. Together with the vanishing of the limit $\mathcal{P}_{k,n\ell m}(1) = 0$, this implies Eq.~\eqref{eq:P_integral_decay}. Since
\begin{align}
    \partial_\lambda\mathcal{P}_{2,\ell} &= P_{\ell-1}^{(1,1)}(\lambda)S_{2,1\ell}(\lambda)(1 - \lambda^2)^{-2} \,,
\end{align}
and because $P_{\ell-1}^{(1,1)}$ is continuous and non-vanishing at $\lambda = 1$, we need $S_{2,1\ell m} = O(1-\lambda)^2$. Since $S_{2,1m}$ is quadratic in $h_{ab}$, this will be the case since by assumption $h_{ab} = O(1-\lambda)$ (cf.~\eqref{eq:metric_regularity_scri}). This regularity assumption also guarantees that Eqs.~\eqref{eq:Q_integral_decay}, \eqref{eq:P_integral_decay2}, \eqref{eq:I_integral_decay} and \eqref{eq:J_integral_decay} are satisfied, and implies that the lower limit $-1$ on the integral Eq.~\eqref{eq:P_Greens_integral} exists. 

Of course, a homogeneous solution which peels (and hence is $\mathcal{PT}$-symmetric at the leading order) may be added to the particular solution~\eqref{eq:particular_solution} and~\eqref{eq:particular_solution2} to yield another peeling solution. In summary, with the assumptions on the metric regularity given by Eq.~\eqref{eq:asymptotic-regular-def} and Eq.~\eqref{eq:metric_regularity_scri}, solutions to Einstein's equations for which all $\Psi_k$ except $\Psi_0$ peel at $\scrip$ are given by
\begin{align}
    \label{eq:peeling_solution}
    \Psi_k &= \sum_{n,\ell-|2-k|\geq0,m}R^{-3-n}\Psi_{k,n\ell m}\,{}_{2-k}Y_{\ell m} = \sum_{n,\ell-|2-k|\geq0,m}R^{-3-n}(\psi_{k,n\ell m} + \Upsilon_{k,n\ell m})\;{}_{2-k}Y_{\ell m} \,.
\end{align}
Here, $\psi_{k,n\ell m}$ is given by Eq.~\eqref{eq:Jacobi_sol} with $s=2$ and $\mathcal{B}_{k,0\ell m} = \mathcal{B}_{k,1\ell m} = 0$. The particular solution $\Upsilon_{k,n\ell m}$ is given by Eq.~\eqref{eq:particular_solution} and Eq.~\eqref{eq:particular_solution2}. 

\begin{remark}
    We conjecture that the metric regularity condition Eq.~\eqref{eq:metric_regularity_scri} does not need to be assumed, but follows from the regularity of the Weyl curvature in combination with a good choice of gauge. If this is true, then all conditions in categories \ref{enum:peeling_1} and \ref{enum:peeling_2} are satisfied and, like for the homogeneous case, only a finite number of terms in the $1/R$ expansion can break peeling for $\Psi_{k\neq0}$. Considerable gauge freedom is left, since (in addition to fixing the metric at $\scri$) we have only fixed the $\mathcal{PT}$ anti-symmetric of $h_{ab}$. The remaining gauge freedom may be used, for example, to apply the \textit{Newman-Unti gauge} $h_{ab}l^b = 0$ at $\scrip$ and $h_{ab}n^b = 0$ at $\scrim$, by setting
    \begin{align}
        (1+f)h_{ab}l^b - (1-f)h_{ab}n^b &= 0 \,,
    \end{align}
    where $f$ is a transition function, antisymmetric about $\lambda = 0$ taking on the values $f(\lambda > \epsilon) = 1$ and $f(\lambda < -\epsilon)=-1$ for some $\epsilon < 1$, so that $\mathcal{PT}(1\pm f) = 1\mp f$. That this gauge is compatible with the gauge $\mathcal{PT}h_{ab} = h_{ab}$ can be easily checked: 
    \begin{align}
        \mathcal{PT}[(1+f)h_{ab}l^b - (1-f)h_{ab}n^b] &= (1-f)\mathcal{PT}[h_{ab}](-n^b) - (1+f)\mathcal{PT}[h_{ab}](-l^b) \nonumber \\
        &= (1+f)(h_{ab} + \Delta h_{ab})l^b - (1-f)(h_{ab} + \Delta h_{ab})n^b \,.
\end{align}
\end{remark}

\subsection{Peeling for asymptotically regular space-times}

To end this section, we give the asymptotic form of the solution in Bondi coordinates. With the metric regularity assumptions given by Eqs. \eqref{eq:asymptotic-regular-def} and \eqref{eq:metric_regularity_scri}, The Weyl scalars fall off as 
\begin{subequations}
\begin{align}
    \Psi_4 &= \Psi_4^\circ r^{-1} + o(r^{-1}) \,, \\
    \Psi_3 &= \Psi_3^\circ r^{-2} + o(r^{-2}) \,, \\
    \Psi_2 &= \Psi_2^\circ r^{-3} + o(r^{-3}) & &\textrm{if $\Psi_{k,0}$ is $\mathcal{PT}$-symmetric} , \\
    \Psi_1 &= \Psi_1^\circ r^{-4} + o(r^{-4}) & &\textrm{if $\Psi_{k,0}$ is $\mathcal{PT}$-symmetric and $\Psi_{k,1}$ is $\mathcal{PT}$-antisymmetric} , \\
    \Psi_0 &= O(r^{-5}\log r) \,. 
\end{align}
\end{subequations}
For initial data that is $\mathcal{PT}$-symmetric at leading order in $1/R$, and $\mathcal{PT}$-antisymmetric at subleading order in $1/R$, the leading order Weyl scalars at $\scrip$ are given by 
\begin{align}
    \Psi_k^\circ &= \psi_k^\circ + \Upsilon_k^\circ = \psi_k^\circ + O(|u|^{1-k}) \,,
\end{align}
where the homogeneous part $\psi_k^\circ$ is given by Eq.~\eqref{eq:leading_psi:asymptotic_form}. 

\subsection{Antipodal mapping of charge aspects}

Here, we explore the consequences of the $\mathcal{PT}$-symmetries on the energy and angular momentum charge aspects. In particular, we explore the connection between the charge aspects in the far future of past null infinity to the charge aspects in the far past of future null infinity. This has been explored in other works~\cite{Ashtekar:1979xeo,Ashtekar:2023wfn, Ashtekar:2023zul, kpis} based on different approaches. Here, we study the implications of the $\mathcal{PT}$-symmetry required for peeling  on this matching. In this section we assume that $\Psi_{k,0}$ is $\mathcal{PT}$-symmetric and $\Psi_{k,1}$ is $\mathcal{PT}$-antisymmetric so that both $\Psi_2$ and $\Psi_1$ peel at $\scrip$, and both $\Psi_2$ and $\Psi_3$ peel at $\scrim$. This is important for a well-defined energy and angular momentum aspect at both $\scrip$ and $\scrim$. 

On $\scrip$ the mass aspect behaves as $\Psi_{2}^\circ = \mathcal{A}_{0\ell m}Y_{\ell m} + O(|u|^{-1})$, so that in the limit to $\inot$, the Bondi energy is given by $E_B = -\frac{\mathcal{A}_{000}}{\sqrt{4\pi}G}$. Since to leading order in $1/R$ (which on $\scrip$ is the same as to leading order in $1/|u|$) $\Psi_2^\circ$ is $\mathcal{PT}$-symmetric, this limit will be the same as the limit from $\scrim$. Explicitly, using the fact that the limits $\lim_{\lambda\to\pm1}r^3\Psi_2$ and $\lim_{R\to\infty}r^3\Psi_2$ exist, we can compute: 
\begin{align}
\begin{split}
    \lim_{\to\inot}-\sqrt{4\pi}G E_{B,\scrip} &= \lim_{R\to\infty}\lim_{\lambda\to1}r^3\Psi_{2,\ell=0} = \lim_{\lambda\to1}\lim_{R\to\infty}r^3\Psi_{2,\ell=0} \\
    &= \lim_{\lambda\to1}\frac{\Psi_{2,0\ell=0}(\lambda)}{(1-\lambda^2)^3} = \lim_{\lambda\to-1}\frac{\Psi_{2,0\ell=0}(-\lambda)}{(1-(-\lambda)^2)^3} \\
    &= \lim_{\lambda\to-1}\frac{\mathcal{PT}[\Psi_{2,0\ell=0}](\lambda)}{(1-\lambda^2)^3} = \lim_{\lambda\to-1}\frac{\Psi_{2,0\ell=0}(\lambda)}{(1-\lambda^2)^3} = \lim_{\to\inot}-\sqrt{4\pi}GE_{B,\scrim} \,, 
\end{split}
\end{align}
where $E_{B,\scripm}$ is the Bondi energy at $\scripm$. 

There exists a similar mapping between the angular momentum aspects on $\scrip$ and $\scrim$. Let us assume we are in a frame at $\scrip$ that is shear-free in the limit to $\inot$. Since, to leading order, the space-time is $\mathcal{PT}$-symmetric, we are simultaneously in a frame at $\scrim$ that is shear-free in the limit to $\inot$. In the limit to $\inot$, the angular momentum aspects are then given by $\Psi_{1,\ell=1}^\circ$ on $\scrip$ and $\Psi_{3,\ell=1}^\circ$ on $\scrim$. The limits $\lim_{R\to\infty}r^4\Upsilon_1$ and $\lim_{R\to\infty}r^4\Upsilon_3$ exist, since at order in $1/R^3$ the wave equation for $\Psi_k$ is just the flat homogeneous spin-$2$ wave equation (cf. Eq.~\eqref{eq:leading_order_EFE}). Hence, for the particular solution we can compute:
\begin{align}
\begin{split}
    \lim_{R\to\infty}\lim_{\lambda\to1}r^4(\mathring{\eth}'\Upsilon_1)_{\ell=1} &= \lim_{\lambda\to1}\frac{\mathring{\eth}'[\Upsilon_{1,11m}(\lambda)\;_1Y_{1m}(\theta,\phi)]}{(1-\lambda^2)^4} \\
    &= \lim_{\lambda\to1}\frac{-\mathring{\eth}'\mathcal{PT}[\Upsilon_{3,11m}\;_{-1}Y_{1m}](\lambda,\theta,\phi)}{(1-\lambda^2)^4} \\
    &= \lim_{\lambda\to-1}\frac{-\mathring{\eth}'[\Upsilon_{3,11m}(\lambda)\;_{-1}Y_{1m}(\pi-\theta,\pi+\phi)]}{(1-\lambda^2)^4} \\
    &= \lim_{\lambda\to-1}\frac{-\sqrt{2}\Upsilon_{3,11m}(\lambda)Y_{1m}(\theta,\phi)}{(1-\lambda^2)^4} = \lim_{R\to\infty}\lim_{\lambda\to-1}-r^4(\mathring{\eth}\Upsilon_3)_{\ell=1} , 
\end{split}
\end{align}
where $\mathring{\eth}$ is the $\eth$ operator on the unit two-sphere. We have the explicit homogeneous solution $\psi_1$ (cf. Eq.~\eqref{eq:leading_psi:asymptotic_form}) so we can evaluate the limits of the angular momentum aspects $\psi_{1,\ell=1}$ on $\scripm$ directly: 
\begin{align}
    \lim_{R\to\infty}\lim_{\lambda\to1}r^4\psi_{1,\ell=1} &= \lim_{R\to\infty}\Bigl[-\tfrac{1}{2}\mathcal{A}_{1,01m}R + \mathcal{A}_{1,11m} + O\Bigl(\frac{1}{R}\Bigr)\Bigr] .
\end{align}
Unless $\mathcal{A}_{k,01m} = 0$ (i.e. if the three-momentum vanishes in the limit to $\inot$) the limits $\lim_{R\to\infty}r^4\psi_1$ and $\lim_{R\to\infty}r^4\psi_3$ do not exist. If the three-momentum vanishes then
\begin{align}
    \lim_{\to\inot}r^4(\mathring{\eth}'\Psi_{1})_{\ell=1}|_{\scrip} &= \lim_{\to\inot}-r^4(\mathring{\eth}\Psi_{3})_{\ell=1}|_{\scrim} .
\end{align}

\section{Discussion}
\label{sec:discussion}

In this work, we used an asymptotic expansion of the space-time in a neighborhood of spatial infinity to study the behaviour of the solutions near past and future null infinity $\scrip$ and $\scrim$. We used the Friedrich coordinates $(\lambda,R)$  defined in Eq.~\eqref{eq:Rlambda_def}. This allows us to solve Einstein's equations perturbatively for large $R$, such that the solutions approximately solve Einstein's equations at spatial infinity, in the far past of future null infinity and the far future of past null infinity.

A key input in solving the equations is the class of data that is considered on a spatial slice. In this work we assumed that the space-time is \emph{asymptotically regular}, which is defined by metrics that satisfy Eq.~\eqref{eq:asymptotic-regular-def} and~\eqref{eq:metric_regularity_scri}. This is a restriction on the class of initial data being considered here. In particular, this excludes initial data with terms logarithmic in $r$ that cannot be removed by a change of coordinates. This restriction was considered for simplicity, and can be relaxed to include a broader class of spacetimes in future work.

Within the class of asymptotically regular space-times, $\Psi_2$ falls off at the usual $1/r^3$ peeling rate at $\scri$ if the space-time arises from initial data that is $\mathcal{PT}$-symmetric to leading order in $1/R$. If, in addition, the subleading data is $\mathcal{PT}$-antisymmetric, then $\Psi_1$ falls off at the usual $1/r^4$ peeling rate at $\scrip$. As a simple but important consequence the fall-off rate $\Psi_1 \sim 1/r^4$ guarantees the existence of a well-defined angular momentum at $\scrip$. Another important consequence is that there exists an antipodal mapping between the mass- and angular momentum aspects on $\scrip$ and $\scrim$ at $\inot$, which leads to a matching of the mass and angular momentum from $\scrim$ to $\scrip$ along the lines of~\cite{Strominger:2013jfa}. 

Let us now remark on how restrictive the asymptotic regularity assumption is. Clearly, there exists a very large class of space-times which have such initial data, however we will focus here on astrophysically relevant space-times. First, we note that Eq.~\eqref{eq:metric_regularity_scri} is satisfied for any isolated astrophysical space-time, and is therefore a very weak requirement. Second, note that the initial data for space-times described in numerical relativity typically satisfy Eq.~\eqref{eq:asymptotic-regular-def}. Therefore, the results in this paper can be directly applied to such simulations. However, PN and PM space-times typically have logarithmic terms in their metrics, violating Eq.~\eqref{eq:asymptotic-regular-def}. Nonetheless, similar to the analysis at null infinity for PN systems~\cite{Blanchet:2020ngx, Blanchet:2023pce}, these logarithmic terms could be eliminated by a suitable coordinate transformation for the PN system. 

For scattering of black holes, however, logarithmic tails in the angular momentum aspect at $\scrip$~\cite{Laddha:2018myi} are a clear signature that the logarithmic terms in the initial data cannot be absorbed into a coordinate transformation. This is because if the data at $t=0$ can be expanded in $1/r$, then our analysis shows that fields at null infinity admit a Taylor expansion in $1/u$.\footnote{For PN systems, to obtain a regular spatial infinity, it is necessary to impose stationarity before some early time $-\tau$ as is usually done in a PN expansion~\cite{Blanchet:2013haa}. Thus, it cannot contain such logarithmic tail terms.}
As such, a further analysis is needed to study the effects of these logarithmic terms to such space-times. 

Finally, let us comment on some broader implications and applications of our work. Firstly, it is observationally relevant to know if peeling is violated for astrophysical space-times. A violation of peeling will affect the gravitational radiation emitted by such systems which could be measurable. It also affects the boundary conditions that should be used in numerical simulations of such space-times. Consequently, understanding the extent of peeling in such space-times is crucial.

Furthermore, the formalism introduced here can have applications beyond the study of peeling. For instance, it can be useful for constructing initial data for numerical simulations using Cauchy-Characteristic-Matching (CCM)\cite{Ma:2024hzq}, Cauchy-Characteristic-Extraction (CCE)~\cite{Bishop:1996gt} or hyperboloidal slicing~\cite{Alvares:2025pbi}. For instance, the no incoming radiation condition can be imposed on $\scrim$, and used to construct data on the initial null cone for the characteristic evolution.

\begin{acknowledgements}
We thank Eric Poisson, Juan Valiente Kroon, and Sizeng Ma for fruitful discussions. This work was supported by the Natural Sciences and Engineering Research Council of Canada. N.K. was also supported by the Tsinghua Shui Mu Scholarship.  
\end{acknowledgements}
\appendix

\section{Jacobi Equation and Polynomials}\label{appendix:Jacobi}
The Jacobi differential equation is given by
\begin{equation}
\label{eq:Jacobi-eq}
(1-x^2) y''(x) + \left(\beta - \alpha - (\alpha + \beta +2)x\right) y'(x) + \nu(\nu + \alpha + \beta + 1)y(x)\,,
\end{equation}
and is equivalent to the hypergeometric equation with a change of variables. The equation is symmetric under $\nu\to -\nu-\alpha - \beta -1$, therefore without loss of generality we can assume that $2\nu \geq -\alpha-\beta -1$. Frobenius theory gives us two independent solutions to~\eqref{eq:Jacobi-eq}, with many special cases depending on the values of $\alpha, \beta, \nu$ \cite{bateman_bateman}. In the main text we always have integer $\alpha, \beta, \nu$ with the conditions that $\alpha+\nu\geq0$, $\beta+\nu\geq 0$ and $\alpha+\beta\geq0$. For such parameters, the Frobenius solutions can be split into 2 types, based on the sign of $\nu$, which we describe below. See~\cite{bateman_bateman} for an extensive treatment of all the cases.

\subsection*{Case 1:  Non-negative integer $\nu$}
The first solution of Eq.~\eqref{eq:Jacobi-eq} in the case where $\nu \geq 0$ are the Jacobi polynomials $P_\nu^{(\alpha, \beta)}(x)$, which are polynomials or order $\nu$. If $\alpha>-1$ and $ \beta>-1$ they are orthogonal on the interval $[-1,1]$ with respect to the weight $(1-x)^\alpha(1+x)^\beta$. However, they are defined for all $\alpha$ and $\beta$. Explicitly, the Jacobi polynomials are given by the Rodriguez formula
\begin{equation}
\label{eq:Rodriguez}
P_{\nu}^{(\alpha, \beta)}(x) = \frac{(-1)^\nu}{2^\nu \nu!}(1-x)^{-\alpha}(1+x)^{-\beta} \frac{d^\nu}{dx^\nu}\left((1-x)^{\alpha + \nu}(1+x)^{\beta+\nu}\right)\,,
\end{equation}
and satisfy the recurrence relation
\begin{align}
\label{eq:Jacobi-recurrence}
2\nu&(\nu+\alpha +\beta )(2\nu+\alpha +\beta -2)P_{\nu}^{(\alpha ,\beta )}(x)\nonumber \\
&=(2\nu+\alpha +\beta -1)\left[(2\nu+\alpha +\beta )(2\nu+\alpha +\beta -2)x+\alpha^{2}-\beta^{2}\right] P_{\nu-1}^{(\alpha ,\beta )}(x) \nonumber\\
&-2(\nu+\alpha -1)(\nu+\beta -1)(2\nu+\alpha +\beta )P_{\nu-2}^{(\alpha ,\beta )}(x)\,.
\end{align}
The second solution to Eq.~\eqref{eq:Jacobi-eq} can be given by the Jacobi function of the second kind, $Q_\nu^{(\alpha, \beta)}(x)$. Let us now describe the structure of this solution. We first analyse the case for $\alpha, \beta > -1$, returning to the other cases later. If $\alpha,\beta > -1$, the Jacobi function of the second kind can be expressed as the integral
\begin{equation}
\label{eq:JacobiQ1}
Q_\nu^{(\alpha, \beta)}(x)= \frac{1}{2(x-1)^{\alpha}(1+x)^\beta} \int_{-1}^1 \frac{(1-t)^\alpha (1+t)^\beta P_{\nu}^{(\alpha, \beta)}(t)}{x-t} \, \mathrm{d}t\,.
\end{equation}
By adding and subtracting $P_\nu^{(\alpha, \beta)}(x)$ in the integrand, we get
\begin{equation}
\label{eq:Q-sec}
Q_\nu^{(\alpha, \beta)}(x)= P_\nu^{(\alpha, \beta)}(x) Q_0^{(\alpha, \beta)}(x) - \frac{1}{2(x-1)^\alpha(1+x)^\beta} q_\nu^{(\alpha, \beta)}(x)\,,
\end{equation}
where  $q_\nu^{(\alpha, \beta)}(x)$ are the secondary Jacobi polynomials given by
\begin{equation}
\label{eq:seconday-jacobi}
q_\nu^{(\alpha, \beta)}(x) = \int_{-1}^1 \frac{P_\nu^{(\alpha, \beta)}(x) - P_\nu^{(\alpha, \beta)}(t)}{x-t} (1-t)^\alpha (1+t)^\beta \,\mathrm{d}t\,.
\end{equation}
It can be shown that $q_\nu^{(\alpha, \beta)}(x)$ are $\nu-1$ order polynomials that satisfy the same recurrence relation~\eqref{eq:Jacobi-recurrence}  as the Jacobi polynomials, but with different initial conditions. Namely,
\begin{align}
q_{0}^{(\alpha, \beta)}(x) &= 0\,,\\
q_{1}^{(\alpha, \beta)}(x) &= \frac{1}{2}(2+\alpha+\beta)2^{\alpha+\beta+1}\frac{\Gamma(\alpha+1)\Gamma(\beta+1)}{\Gamma(\alpha+\beta+2)}\,,
\end{align}
which follows from Eq.~\eqref{eq:seconday-jacobi}. Thus, the structure of $Q_\nu^{(\alpha, \beta)}$ can be ascertained from an expression for $Q_0^{(\alpha,\beta)}$. To obtain an explicit equation we use the equation
\begin{align}
    \label{eq:JacobiQ-derivative}
    Q_0^{(\alpha, \beta) \,\prime}(x) &= -\frac{2^{\alpha+\beta}\Gamma(\alpha +1)\Gamma(\beta +1) }{\Gamma(\alpha +\beta +1)(x -1)^{\alpha +1}(x +1)^{\beta+1}}  \,,
\end{align}
which can be derived from the Wronskian of the Jacobi equation.
This can be integrated using Feynman's technique, by introducing parameters $r$ and $t$, and then writing the right-hand side of~\eqref{eq:JacobiQ-derivative} as a term proportional to the derivatives of  $(x-t)^{-1}(x+r)^{-1}$ with respect to $r$ and $t$, evaluated at $r=t=1$. This expression can be easily integrated, by commuting the integral with the derivatives, to give us the expression
\begin{equation}
    \label{eq:Q0}
    Q_0^{(\alpha, \beta)}(x) = \frac{(-1)^{\beta+1 } 2^{\alpha+\beta} }{\Gamma(\alpha +\beta +1)} \frac{d^\alpha}{d t^\alpha} \frac{d^\beta}{dr^\beta}\left(\frac{\log(\frac{t-x}{r+x})}{r+t}\right)\bigg|_{r=1,t=1}\,,
\end{equation}
where the constant of integration can be verified to be 0. From this expression it can be easily verified that $Q_0^{(\alpha, \beta)}(x)$ is of the form
\begin{equation}
\label{eq:Q0-2}
    Q_0^{(\alpha, \beta)}(x) = \frac{(-1)^\alpha}{2}\log\left(\frac{1-x}{1+x}\right) + \frac{p^{(\alpha,\beta)}(x)}{(1-x)^\alpha(1+x)^\beta}\,,
\end{equation}
where $p^{(\alpha,\beta)}(x)$ is some polynomial that can be computed using~\eqref{eq:Q0}. Thus for non-negative integers $\alpha$ and $\beta$, we can use~\eqref{eq:Q-sec} and~\eqref{eq:Q0-2} to obtain the analytic structure of the secondary solutions. 

Now we describe the structure for $\alpha<0, \beta>0$ and $\beta<0,\alpha>0$. We obtain this by using the `raising' and `lowering' operators
\begin{subequations}
\begin{align}
    (1-x)\frac{d}{dx}P^{(\alpha,\beta)}_{\nu}(x) - \alpha P^{(\alpha,\beta)}_{\nu}(x) &= -(\alpha+\nu)P^{(\alpha-1,\beta+1)}_{\nu}(x) \label{eq:alpha-down-P}\\
    (1-x)\frac{d}{dx}Q^{(\alpha,\beta)}_{\nu}(x) - \alpha Q^{(\alpha,\beta)}_{\nu}(x) &= (\alpha+\nu)Q^{(\alpha-1,\beta+1)}_{\nu}(x) \label{eq:alpha-down-Q}\\
    (1+x)\frac{d}{dx}P^{(\alpha,\beta)}_{\nu}(x) + \beta P^{(\alpha,\beta)}_{\nu}(x) &= (\beta+\nu)P^{(\alpha+1,\beta-1)}_{\nu}(x)\\
    (1+x)\frac{d}{dx}Q^{(\alpha,\beta)}_{\nu}(x) + \beta Q^{(\alpha,\beta)}_{\nu}(x) &= -(\beta+\nu)Q^{(\alpha+1,\beta-1)}_{\nu}(x)
\end{align}

First, we analyse the $\alpha<0, \beta>0$ case. Recall that we assume $\alpha+\beta>0$. By the previous analysis we know that the $Q^{(0,\alpha+\beta)}_\nu(x)$ is of the form
\begin{equation}
    \label{eq:Q0aplusb}
    Q^{(0,\alpha+\beta)}_\nu(x) = (-1)^\alpha P_\nu^{(0,\alpha+\beta)}(x)\log\left(\frac{1-x}{1+x}\right) + \frac{\tilde{p}^{(0,\alpha+\beta)}_\nu (x)}{(1+x)^{\alpha+\beta}}\,,
\end{equation}
where $\tilde{p}^{(0,\alpha+\beta)}$ is a polynomial. Now, the application of the operator $(1-x)\tfrac{d}{dx} - \alpha$ to~\eqref{eq:Q0aplusb} $|\alpha|$ times gives us
\begin{equation}
    Q^{(\alpha, \beta)}_\nu(x) = (-1)^\alpha P_\nu^{(\alpha,\beta)}(x)\log\left(\frac{1-x}{1+x}\right) + \frac{\tilde{p}^{(\alpha,\beta)}_\nu (x)}{(1+x)^{\beta}}\,,
\end{equation}
where we have used~\eqref{eq:alpha-down-P} and~\eqref{eq:alpha-down-Q} at each step. Similarly, it is easy to show that for $\alpha>0,\beta<0$, we get $Q^{(\alpha,\beta)}_{\nu}(x)$ of the form
\begin{equation}
    Q^{(\alpha, \beta)}_\nu(x) = (-1)^\alpha P_\nu^{(\alpha,\beta)}(x)\log\left(\frac{1-x}{1+x}\right) + \frac{\tilde{p}^{(\alpha,\beta)}_\nu (x)}{(1-x)^{\alpha}}\,,
\end{equation}

\end{subequations}
\subsubsection{Asymptotics}
The asymptotics of the Jacobi Polynomials as $x\to \pm1$ can be easily seen by expanding the Rodriguez formula~\eqref{eq:Rodriguez}. If $\nu \geq 0$ and either $\alpha \geq 0$, or $\alpha + \nu < 0$ the Jacobi polynomials have non-zero values at $x = 1$ given by
\begin{equation}
P_\nu ^{(\alpha, \beta)}(1) = \frac{(\alpha+1)_\nu}{\nu!} \,.
\end{equation}
Similarly, if $\nu \geq 0$ and either $\beta \geq 0$ or $\beta + \nu < 0$, the Jacobi polynomials have non-zero values at $x = -1$ given by
\begin{equation}
P_\nu ^{(\alpha, \beta)}(-1) = \frac{(\beta+1)_\nu}{\nu!}\,.
\end{equation}
If, on the other hand, $\alpha < 0$ and $\alpha + \nu \geq 0$, the asymptotics of the Jacobi Polynomial as $x$ approaches $1$ are
\begin{equation}
P_\nu ^{(\alpha, \beta)}(x) \sim (1-x)^{-\alpha}+ O\left((1-x)^{-\alpha+1} \right)\,.
\end{equation}
Similarly, if $\beta\leq0$ and $\beta+\nu \geq 0$, the asymptotics of the Jacobi Polynomial as $x$ approaches $-1$ are
\begin{equation}
P_\nu ^{(\alpha, \beta)}(x) \sim (1+x)^{-\beta}+ O\left((1+x)^{-\beta+1} \right)\,.
\end{equation}

For the Jacobi function of the second kind $Q_{\nu}^{(\alpha,\beta)}(x)$, note that the coefficient of the $\log$ term is the Jacobi polynomial $P_{\nu}^{(\alpha,\beta)}(x)$. Therefore, the asymptotics of the Jacobi polynomial above can be used to obtain its asymptotics. Furthermore, the non-log terms go as $(1-x)^{-\textrm{max}(0,\alpha)}$ for $x\to1$ and $(1+x)^{-\textrm{max}(0,\beta)}$ for $x\to-1$.

\subsection{Case 2: negative integer $\nu$}
In this case $\nu<0$ is a negative integer and $\nu > - \alpha - \beta - 1$, so $\nu$ cannot be transformed to a positive number using the symmetries of the Jacobi equation. The Jacobi equation then has the following 2 linearly independent solutions:
\begin{equation}
\label{eq:neg-nu-first}
\frac{P^{(\alpha, -\beta)}_{\beta+\nu}(x)}{(1+x)^\beta}\,,
\end{equation}
and
\begin{equation}
\label{eq:neg-nu-second}
\frac{P^{(-\alpha, \beta)}_{\alpha+\nu}(x)}{(1-x)^\alpha}\,.
\end{equation}
These are both rational functions of $x$. Using the asymptotics of the Jacobi Polynomial above, we can obtain the asymptotics of the solutions. First, note that by assumption $\alpha + \nu \geq 0$ and $\beta + \nu \geq 0$, implying in particular that $\alpha>0$ and $\beta>0$. Thus, this tells use that that the first solution~\eqref{eq:neg-nu-first} goes as $(1+x)^{-\beta}$ as $x\to-1$  and $(1-x)^{0}$ for $x\to1$. Similarly, for the second solution we get $(1+x)^{0}$ for $x\to-1$  and $(1-x)^{-\alpha}$ for $x\to1$.

\section{A GHP solution to the Penrose wave equation}
\label{appendix:gen_sol}
Here, we will derive a GHP solution to the spin-$s$ wave equation \eqref{eq:GHP_wave_a}. As a starting point, let us first consider the lowest $\ell = s - k$ mode of $\phi_k$, which satisfies $\eth'\eth\phi_{k,\ell m} = 0$ if $\ell = s - k \geq 0$, and $\eth\eth'\phi_{k,\ell m} = 0$ if $\ell = s - k \leq 0$. It follows that these modes satisfy
\begin{subequations}
\begin{align}
    \th_c'\th_c\phi_{k,\ell m} &= 0 & &\textrm{if} \; \ell = s - k \geq 0 \,, \\
    \th_c\th_c'\phi_{k,\ell m} &= 0 & &\textrm{if} \; \ell = s - k \leq 0 \,. 
\end{align}
\end{subequations}
GHP solutions to these equations satisfying
\begin{subequations}
\begin{align}
    \th_c\phi_{k,\ell m} &= (\th - [2s-k+1]\rho)\phi_{k,\ell m} = 0 & &\textrm{if} \; \ell = s - k \geq 0 \,, \\
    \th_c'\phi_{k,\ell m} &= (\th' - [k+1]\rho')\phi_{k,\ell m} = 0 & &\textrm{if} \; \ell = s - k \leq 0 \,, 
\end{align}
\end{subequations}
are given by
\begin{subequations}
\begin{align}
    \phi_{k,\ell m} &= \rho^{2s-k+1}U_{\ell m}(u){}_{s-k}Y_{\ell m} & &\textrm{if} \; \ell = s - k \geq 0 \,, \\
    \phi_{k,\ell m} &= \rho^{k+1}V_{\ell m}(v){}_{k-s}Y_{\ell m} & &\textrm{if} \; \ell = s - k \leq 0 \,, 
\end{align}
\end{subequations}
where $U$ and $V$ are arbitrary functions, which can be easily seen from the fact that $\th\rho = \rho^2$ and $\th'\rho = \rho\rho'$. 

The operators $\eth'^\dagger\th_c$ and $\eth^\dagger\th_c'$, defined in Eq.~\eqref{eq:ladder} of the main text, can be used to derive higher mode solutions to the spin-$s$ wave equation \eqref{eq:GHP_wave_a}. These operators have the important property that they commute with the wave operator $\Ro$. To establish this, we first need the important commutators
\begin{subequations}
\begin{align}
    \Ro - \Ro' &= 0 \\
    [\th_c, \eth] &= 0 \\
    [\th_c, \eth^\dagger] &= 0 \,.
\end{align}
\end{subequations}
The first two of these easily follow from \eqref{eq:conformal_commutators}. The last can be seen by first noting that the radial dependence of $\eth^\dagger$ is $\eth^\dagger = r\mathring{\eth}^\dagger = A\rho^{-1}\mathring{\eth}^\dagger$ for some (weighted) constant $A$ (cf.~\eqref{eq:eth_dagger}), and where $\mathring{\eth}^\dagger$ is a purely angular operator. Hence, since $\th r = A\th\rho^{-1} = -A = -r\rho$, 
\begin{align}
    \th_c\eth^\dagger\eta &= (\th - [1 + s + \tfrac{1}{2}(p-q+2)]\rho)(r\mathring{\eth}^\dagger\eta) = r\mathring{\eth}^\dagger(\th - [1 + s + \tfrac{1}{2}(p-q)]\rho)\eta = \eth^\dagger\th_c\eta \,.
\end{align}
We can now compute
\begin{align}
    [\Ro, \eth^\dagger\th'_c] &= (\th_c\th'_c - \eth\eth')\eth^\dagger\th'_c - \eth^\dagger\th'_c(\th_c\th'_c - \eth\eth') \nonumber \\
    &= (\th_c\th'_c - \eth\eth')\eth^\dagger\th'_c - (\th'_c\th_c - \eth'\eth)\eth^\dagger\th'_c \\
    &= (\Ro - \Ro')\eth^\dagger\th'_c = 0 \,. \nonumber
\end{align}
Since $\eth'^\dagger\th_c$ and $\eth^\dagger\th_c$ raise and lower spin respectively, if $s - k \geq 0$ it follows that the $\ell = s - k + 1$ mode of $\phi_k$ satisfies
\begin{align}
    \th_c'\th_c\eth'^\dagger\th_c\phi_{k,\ell=s-k+1} &= 0 \,,
\end{align}
which has a GHP solution
\begin{align}
\label{eq:ladder_phi}
    \eth'^\dagger\th_c\phi_{k,\ell=s-k+1} &= \rho^{2s-k+2}U_{\ell=s-k+1}(u){}_{s-k+1}Y_{\ell=s-k+1} \,.
\end{align}
$\phi_{k,\ell=s-k+1}$ can be recovered by applying $\eth^\dagger\th_c'$ to Eq.~\eqref{eq:ladder_phi}, since
\begin{align}
    \eth^\dagger\th_c'\eth'^\dagger\th_c\phi_{k,\ell=s-k+1} = \eth^\dagger\eth'^\dagger\th_c'\th_c\phi_{k,\ell=s-k+1} = \eth^\dagger\eth'^\dagger\eth'\eth\phi_{k,\ell=s-k+1} = \phi_{k,\ell=s-k+1} \,.
\end{align}
Generalizing, we find that for any $\ell$-mode, a GHP solution to $\Ro\phi_{k,\ell} = 0$ is given by
\begin{align}
\label{eq:GHP_sol_1}
    \phi_{k,\ell m} &= (\eth^\dagger\th_c')^{\ell-s+k}(\rho^{\ell+s+1}U_{\ell m}(u){}_\ell Y_{\ell m}) \,,
\end{align}
where without loss of generality we assume $s - k \geq 0$. A second solution can be obtained by noting that, by the massless free-field equations \eqref{eq:massless_f}, $\phi_k = (\eth'^\dagger\th_c)^{2(s-k)}\phi_{2s-k}$, where we choose the solution
\begin{align}
    \phi_{2s-k,\ell m} &= (\eth'^\dagger\th_c)^{\ell-s+k}(\rho^{\ell+s+1}V_{\ell m}(v){}_{-\ell}Y_{\ell m}) \,,
\end{align}
so that
\begin{align}
\label{eq:GHP_sol_2}
    \phi_{k,\ell m} &= (\eth'^\dagger\th_c)^{\ell+s-k}(\rho^{\ell+s+1}V_{\ell m}(v){}_{-\ell}Y_{\ell m}) \,.
\end{align}
The general solution will be a linear combination of the two solutions \eqref{eq:GHP_sol_1} and \eqref{eq:GHP_sol_2}: 
\begin{align}
    \phi_{k,\ell m} &= (\eth^\dagger\th_c')^{\ell-s+k}(\rho^{\ell+s+1}U_{\ell m}(u){}_\ell Y_{\ell m}) + (\eth'^\dagger\th_c)^{\ell+s-k}(\rho^{\ell+s+1}V_{\ell m}(v){}_{-\ell}Y_{\ell m}) \nonumber \\
    &= \bigl[\th_c'^{\ell-s+k}(\rho^{2s-k+1}\tilde{U}_{\ell m}(u)) + \th_c^{\ell+s-k}(\rho^{k+1}\tilde{V}_{\ell m}(v))\bigr]{}_{s-k}Y_{\ell m} \,,
\end{align}
where $U$ and $\tilde{U}$, and $V$ and $\tilde{V}$ differ by a numerical factor. 
\bibliography{references}

\begin{thebibliography}{44}%
\makeatletter
\providecommand \@ifxundefined [1]{%
 \@ifx{#1\undefined}
}%
\providecommand \@ifnum [1]{%
 \ifnum #1\expandafter \@firstoftwo
 \else \expandafter \@secondoftwo
 \fi
}%
\providecommand \@ifx [1]{%
 \ifx #1\expandafter \@firstoftwo
 \else \expandafter \@secondoftwo
 \fi
}%
\providecommand \natexlab [1]{#1}%
\providecommand \enquote  [1]{``#1''}%
\providecommand \bibnamefont  [1]{#1}%
\providecommand \bibfnamefont [1]{#1}%
\providecommand \citenamefont [1]{#1}%
\providecommand \href@noop [0]{\@secondoftwo}%
\providecommand \href [0]{\begingroup \@sanitize@url \@href}%
\providecommand \@href[1]{\@@startlink{#1}\@@href}%
\providecommand \@@href[1]{\endgroup#1\@@endlink}%
\providecommand \@sanitize@url [0]{\catcode `\\12\catcode `\$12\catcode `\&12\catcode `\#12\catcode `\^12\catcode `\_12\catcode `\%12\relax}%
\providecommand \@@startlink[1]{}%
\providecommand \@@endlink[0]{}%
\providecommand \url  [0]{\begingroup\@sanitize@url \@url }%
\providecommand \@url [1]{\endgroup\@href {#1}{\urlprefix }}%
\providecommand \urlprefix  [0]{URL }%
\providecommand \Eprint [0]{\href }%
\providecommand \doibase [0]{https://doi.org/}%
\providecommand \selectlanguage [0]{\@gobble}%
\providecommand \bibinfo  [0]{\@secondoftwo}%
\providecommand \bibfield  [0]{\@secondoftwo}%
\providecommand \translation [1]{[#1]}%
\providecommand \BibitemOpen [0]{}%
\providecommand \bibitemStop [0]{}%
\providecommand \bibitemNoStop [0]{.\EOS\space}%
\providecommand \EOS [0]{\spacefactor3000\relax}%
\providecommand \BibitemShut  [1]{\csname bibitem#1\endcsname}%
\let\auto@bib@innerbib\@empty
\bibitem [{\citenamefont {Bondi}\ \emph {et~al.}(1962)\citenamefont {Bondi}, \citenamefont {van~der Burg},\ and\ \citenamefont {Metzner}}]{bondi}%
  \BibitemOpen
  \bibfield  {author} {\bibinfo {author} {\bibfnamefont {H.}~\bibnamefont {Bondi}}, \bibinfo {author} {\bibfnamefont {M.~G.~J.}\ \bibnamefont {van~der Burg}},\ and\ \bibinfo {author} {\bibfnamefont {A.}~\bibnamefont {Metzner}},\ }\bibfield  {title} {\bibinfo {title} {{Gravitational waves in general relativity, VII. Waves from axi-symmetric isolated system}},\ }\href@noop {} {\bibfield  {journal} {\bibinfo  {journal} {Proceedings of the Royal Society of London. Series A. Mathematical and Physical Sciences}\ }\textbf {\bibinfo {volume} {269}},\ \bibinfo {pages} {21} (\bibinfo {year} {1962})}\BibitemShut {NoStop}%
\bibitem [{\citenamefont {Sachs}(1962{\natexlab{a}})}]{sachs}%
  \BibitemOpen
  \bibfield  {author} {\bibinfo {author} {\bibfnamefont {R.~K.}\ \bibnamefont {Sachs}},\ }\bibfield  {title} {\bibinfo {title} {{Gravitational waves in general relativity VIII. Waves in asymptotically flat space-time}},\ }\href@noop {} {\bibfield  {journal} {\bibinfo  {journal} {Proceedings of the Royal Society of London. Series A. Mathematical and Physical Sciences}\ }\textbf {\bibinfo {volume} {270}},\ \bibinfo {pages} {103} (\bibinfo {year} {1962}{\natexlab{a}})}\BibitemShut {NoStop}%
\bibitem [{\citenamefont {Sachs}(1962{\natexlab{b}})}]{sachs2}%
  \BibitemOpen
  \bibfield  {author} {\bibinfo {author} {\bibfnamefont {R.}~\bibnamefont {Sachs}},\ }\bibfield  {title} {\bibinfo {title} {{Asymptotic symmetries in gravitational theory}},\ }\href {https://doi.org/10.1103/PhysRev.128.2851} {\bibfield  {journal} {\bibinfo  {journal} {Phys. Rev.}\ }\textbf {\bibinfo {volume} {128}},\ \bibinfo {pages} {2851} (\bibinfo {year} {1962}{\natexlab{b}})}\BibitemShut {NoStop}%
\bibitem [{\citenamefont {Penrose}(1965)}]{rp}%
  \BibitemOpen
  \bibfield  {author} {\bibinfo {author} {\bibfnamefont {R.}~\bibnamefont {Penrose}},\ }\bibfield  {title} {\bibinfo {title} {Zero rest-mass fields including gravitation: asymptotic behaviour},\ }\href@noop {} {\bibfield  {journal} {\bibinfo  {journal} {Proceedings of the Royal Society of London. Series A. Mathematical and physical sciences}\ }\textbf {\bibinfo {volume} {284}},\ \bibinfo {pages} {159} (\bibinfo {year} {1965})}\BibitemShut {NoStop}%
\bibitem [{\citenamefont {Christodoulou}\ and\ \citenamefont {Klainerman}(1993)}]{dcsk}%
  \BibitemOpen
  \bibfield  {author} {\bibinfo {author} {\bibfnamefont {D.}~\bibnamefont {Christodoulou}}\ and\ \bibinfo {author} {\bibfnamefont {S.}~\bibnamefont {Klainerman}},\ }\bibfield  {title} {\bibinfo {title} {{The global nonlinear stability of the Minkowski space}},\ }\href@noop {} {\bibfield  {journal} {\bibinfo  {journal} {S{\'e}minaire {\'E}quations aux d{\'e}riv{\'e}es partielles (Polytechnique) dit aussi" S{\'e}minaire Goulaouic-Schwartz"}\ ,\ \bibinfo {pages} {1}} (\bibinfo {year} {1993})}\BibitemShut {NoStop}%
\bibitem [{\citenamefont {Bieri}(2010)}]{bieri}%
  \BibitemOpen
  \bibfield  {author} {\bibinfo {author} {\bibfnamefont {L.}~\bibnamefont {Bieri}},\ }\bibfield  {title} {\bibinfo {title} {{An Extension of the Stability Theorem of the Minkowski Space in General Relativity}},\ }\href@noop {} {\bibfield  {journal} {\bibinfo  {journal} {J. Diff. Geom.}\ }\textbf {\bibinfo {volume} {86}},\ \bibinfo {pages} {17} (\bibinfo {year} {2010})},\ \Eprint {https://arxiv.org/abs/0904.0620} {arXiv:0904.0620 [gr-qc]} \BibitemShut {NoStop}%
\bibitem [{\citenamefont {Kehrberger}(2022{\natexlab{a}})}]{Kehrberger:2021uvf}%
  \BibitemOpen
  \bibfield  {author} {\bibinfo {author} {\bibfnamefont {L.~M.~A.}\ \bibnamefont {Kehrberger}},\ }\bibfield  {title} {\bibinfo {title} {{The Case Against Smooth Null Infinity I: Heuristics and Counter-Examples}},\ }\href {https://doi.org/10.1007/s00023-021-01108-2} {\bibfield  {journal} {\bibinfo  {journal} {Annales Henri Poincare}\ }\textbf {\bibinfo {volume} {23}},\ \bibinfo {pages} {829} (\bibinfo {year} {2022}{\natexlab{a}})},\ \Eprint {https://arxiv.org/abs/2105.08079} {arXiv:2105.08079 [gr-qc]} \BibitemShut {NoStop}%
\bibitem [{\citenamefont {Kehrberger}(2021)}]{Kehrberger:2021vhp}%
  \BibitemOpen
  \bibfield  {author} {\bibinfo {author} {\bibfnamefont {L.~M.~A.}\ \bibnamefont {Kehrberger}},\ }\bibfield  {title} {\bibinfo {title} {{The Case Against Smooth Null Infinity II: A Logarithmically Modified Price's Law}},\ }\href@noop {} {\  (\bibinfo {year} {2021})},\ \Eprint {https://arxiv.org/abs/2105.08084} {arXiv:2105.08084 [gr-qc]} \BibitemShut {NoStop}%
\bibitem [{\citenamefont {Kehrberger}(2022{\natexlab{b}})}]{Kehrberger:2021azo}%
  \BibitemOpen
  \bibfield  {author} {\bibinfo {author} {\bibfnamefont {L.~M.~A.}\ \bibnamefont {Kehrberger}},\ }\bibfield  {title} {\bibinfo {title} {{The Case Against Smooth Null Infinity III: Early-Time Asymptotics for Higher $\ell $-Modes of Linear Waves on a Schwarzschild Background}},\ }\href {https://doi.org/10.1007/s40818-022-00129-2} {\bibfield  {journal} {\bibinfo  {journal} {Ann. PDE}\ }\textbf {\bibinfo {volume} {8}},\ \bibinfo {pages} {12} (\bibinfo {year} {2022}{\natexlab{b}})},\ \Eprint {https://arxiv.org/abs/2106.00035} {arXiv:2106.00035 [gr-qc]} \BibitemShut {NoStop}%
\bibitem [{\citenamefont {Kehrberger}(2024)}]{Kehrberger:2024clh}%
  \BibitemOpen
  \bibfield  {author} {\bibinfo {author} {\bibfnamefont {L.}~\bibnamefont {Kehrberger}},\ }\bibfield  {title} {\bibinfo {title} {{The Case Against Smooth Null Infinity IV: Linearized Gravity Around Schwarzschild\textemdash{}An Overview}},\ }\href {https://doi.org/10.1098/rsta.2023.0039} {\bibfield  {journal} {\bibinfo  {journal} {Phil. Trans. Roy. Soc. Lond. A}\ }\textbf {\bibinfo {volume} {382}},\ \bibinfo {pages} {20230039} (\bibinfo {year} {2024})},\ \Eprint {https://arxiv.org/abs/2401.04170} {arXiv:2401.04170 [gr-qc]} \BibitemShut {NoStop}%
\bibitem [{\citenamefont {Kehrberger}\ and\ \citenamefont {Masaood}(2024)}]{Kehrberger:2024aak}%
  \BibitemOpen
  \bibfield  {author} {\bibinfo {author} {\bibfnamefont {L.}~\bibnamefont {Kehrberger}}\ and\ \bibinfo {author} {\bibfnamefont {H.}~\bibnamefont {Masaood}},\ }\bibfield  {title} {\bibinfo {title} {{The Case Against Smooth Null Infinity V: Early-Time Asymptotics of Linearised Gravity Around Schwarzschild for Fixed Spherical Harmonic Modes}},\ }\href@noop {} {\  (\bibinfo {year} {2024})},\ \Eprint {https://arxiv.org/abs/2401.04179} {arXiv:2401.04179 [gr-qc]} \BibitemShut {NoStop}%
\bibitem [{\citenamefont {Bishop}\ \emph {et~al.}(1996)\citenamefont {Bishop}, \citenamefont {Gomez}, \citenamefont {Lehner},\ and\ \citenamefont {Winicour}}]{Bishop:1996gt}%
  \BibitemOpen
  \bibfield  {author} {\bibinfo {author} {\bibfnamefont {N.~T.}\ \bibnamefont {Bishop}}, \bibinfo {author} {\bibfnamefont {R.}~\bibnamefont {Gomez}}, \bibinfo {author} {\bibfnamefont {L.}~\bibnamefont {Lehner}},\ and\ \bibinfo {author} {\bibfnamefont {J.}~\bibnamefont {Winicour}},\ }\bibfield  {title} {\bibinfo {title} {{Cauchy-characteristic extraction in numerical relativity}},\ }\href {https://doi.org/10.1103/PhysRevD.54.6153} {\bibfield  {journal} {\bibinfo  {journal} {Phys. Rev. D}\ }\textbf {\bibinfo {volume} {54}},\ \bibinfo {pages} {6153} (\bibinfo {year} {1996})},\ \Eprint {https://arxiv.org/abs/gr-qc/9705033} {arXiv:gr-qc/9705033} \BibitemShut {NoStop}%
\bibitem [{\citenamefont {Reisswig}\ \emph {et~al.}(2010)\citenamefont {Reisswig}, \citenamefont {Bishop}, \citenamefont {Pollney},\ and\ \citenamefont {Szilagyi}}]{Reisswig:2009rx}%
  \BibitemOpen
  \bibfield  {author} {\bibinfo {author} {\bibfnamefont {C.}~\bibnamefont {Reisswig}}, \bibinfo {author} {\bibfnamefont {N.~T.}\ \bibnamefont {Bishop}}, \bibinfo {author} {\bibfnamefont {D.}~\bibnamefont {Pollney}},\ and\ \bibinfo {author} {\bibfnamefont {B.}~\bibnamefont {Szilagyi}},\ }\bibfield  {title} {\bibinfo {title} {{Characteristic extraction in numerical relativity: binary black hole merger waveforms at null infinity}},\ }\href {https://doi.org/10.1088/0264-9381/27/7/075014} {\bibfield  {journal} {\bibinfo  {journal} {Class. Quant. Grav.}\ }\textbf {\bibinfo {volume} {27}},\ \bibinfo {pages} {075014} (\bibinfo {year} {2010})},\ \Eprint {https://arxiv.org/abs/0912.1285} {arXiv:0912.1285 [gr-qc]} \BibitemShut {NoStop}%
\bibitem [{\citenamefont {Moxon}\ \emph {et~al.}(2023)\citenamefont {Moxon}, \citenamefont {Scheel}, \citenamefont {Teukolsky}, \citenamefont {Deppe}, \citenamefont {Fischer}, \citenamefont {H{\'e}bert}, \citenamefont {Kidder},\ and\ \citenamefont {Throwe}}]{Moxon:2021gbv}%
  \BibitemOpen
  \bibfield  {author} {\bibinfo {author} {\bibfnamefont {J.}~\bibnamefont {Moxon}}, \bibinfo {author} {\bibfnamefont {M.~A.}\ \bibnamefont {Scheel}}, \bibinfo {author} {\bibfnamefont {S.~A.}\ \bibnamefont {Teukolsky}}, \bibinfo {author} {\bibfnamefont {N.}~\bibnamefont {Deppe}}, \bibinfo {author} {\bibfnamefont {N.}~\bibnamefont {Fischer}}, \bibinfo {author} {\bibfnamefont {F.}~\bibnamefont {H{\'e}bert}}, \bibinfo {author} {\bibfnamefont {L.~E.}\ \bibnamefont {Kidder}},\ and\ \bibinfo {author} {\bibfnamefont {W.}~\bibnamefont {Throwe}},\ }\bibfield  {title} {\bibinfo {title} {{SpECTRE Cauchy-characteristic evolution system for rapid, precise waveform extraction}},\ }\href {https://doi.org/10.1103/PhysRevD.107.064013} {\bibfield  {journal} {\bibinfo  {journal} {Phys. Rev. D}\ }\textbf {\bibinfo {volume} {107}},\ \bibinfo {pages} {064013} (\bibinfo {year} {2023})},\ \Eprint {https://arxiv.org/abs/2110.08635} {arXiv:2110.08635 [gr-qc]} \BibitemShut {NoStop}%
\bibitem [{\citenamefont {Iozzo}\ \emph {et~al.}(2021)\citenamefont {Iozzo} \emph {et~al.}}]{Iozzo:2021vnq}%
  \BibitemOpen
  \bibfield  {author} {\bibinfo {author} {\bibfnamefont {D.~A.~B.}\ \bibnamefont {Iozzo}} \emph {et~al.},\ }\bibfield  {title} {\bibinfo {title} {{Comparing Remnant Properties from Horizon Data and Asymptotic Data in Numerical Relativity}},\ }\href {https://doi.org/10.1103/PhysRevD.103.124029} {\bibfield  {journal} {\bibinfo  {journal} {Phys. Rev. D}\ }\textbf {\bibinfo {volume} {103}},\ \bibinfo {pages} {124029} (\bibinfo {year} {2021})},\ \Eprint {https://arxiv.org/abs/2104.07052} {arXiv:2104.07052 [gr-qc]} \BibitemShut {NoStop}%
\bibitem [{\citenamefont {Ma}\ \emph {et~al.}(2024)\citenamefont {Ma}, \citenamefont {Scheel}, \citenamefont {Moxon}, \citenamefont {Nelli}, \citenamefont {Deppe}, \citenamefont {Kidder}, \citenamefont {Throwe},\ and\ \citenamefont {Vu}}]{Ma:2024hzq}%
  \BibitemOpen
  \bibfield  {author} {\bibinfo {author} {\bibfnamefont {S.}~\bibnamefont {Ma}}, \bibinfo {author} {\bibfnamefont {M.~A.}\ \bibnamefont {Scheel}}, \bibinfo {author} {\bibfnamefont {J.}~\bibnamefont {Moxon}}, \bibinfo {author} {\bibfnamefont {K.~C.}\ \bibnamefont {Nelli}}, \bibinfo {author} {\bibfnamefont {N.}~\bibnamefont {Deppe}}, \bibinfo {author} {\bibfnamefont {L.~E.}\ \bibnamefont {Kidder}}, \bibinfo {author} {\bibfnamefont {W.}~\bibnamefont {Throwe}},\ and\ \bibinfo {author} {\bibfnamefont {N.~L.}\ \bibnamefont {Vu}},\ }\bibfield  {title} {\bibinfo {title} {{Merging black holes with Cauchy-characteristic matching: Computation of late-time tails}},\ }\href@noop {} {\  (\bibinfo {year} {2024})},\ \Eprint {https://arxiv.org/abs/2412.06906} {arXiv:2412.06906 [gr-qc]} \BibitemShut {NoStop}%
\bibitem [{\citenamefont {Arnowitt}\ \emph {et~al.}(2008)\citenamefont {Arnowitt}, \citenamefont {Deser},\ and\ \citenamefont {Misner}}]{adm}%
  \BibitemOpen
  \bibfield  {author} {\bibinfo {author} {\bibfnamefont {R.}~\bibnamefont {Arnowitt}}, \bibinfo {author} {\bibfnamefont {S.}~\bibnamefont {Deser}},\ and\ \bibinfo {author} {\bibfnamefont {C.~W.}\ \bibnamefont {Misner}},\ }\bibfield  {title} {\bibinfo {title} {Republication of: The dynamics of general relativity},\ }\href@noop {} {\bibfield  {journal} {\bibinfo  {journal} {General Relativity and Gravitation}\ }\textbf {\bibinfo {volume} {40}},\ \bibinfo {pages} {1997} (\bibinfo {year} {2008})}\BibitemShut {NoStop}%
\bibitem [{\citenamefont {Geroch}(1972)}]{rg-jmp}%
  \BibitemOpen
  \bibfield  {author} {\bibinfo {author} {\bibfnamefont {R.}~\bibnamefont {Geroch}},\ }\bibfield  {title} {\bibinfo {title} {Structure of the gravitational field at spatial infinity},\ }\href@noop {} {\bibfield  {journal} {\bibinfo  {journal} {Journal of Mathematical Physics}\ }\textbf {\bibinfo {volume} {13}},\ \bibinfo {pages} {956} (\bibinfo {year} {1972})}\BibitemShut {NoStop}%
\bibitem [{\citenamefont {Ashtekar}\ and\ \citenamefont {Hansen}(1978)}]{Ashtekar:1978zz}%
  \BibitemOpen
  \bibfield  {author} {\bibinfo {author} {\bibfnamefont {A.}~\bibnamefont {Ashtekar}}\ and\ \bibinfo {author} {\bibfnamefont {R.~O.}\ \bibnamefont {Hansen}},\ }\bibfield  {title} {\bibinfo {title} {{A unified treatment of null and spatial infinity in general relativity. I - Universal structure, asymptotic symmetries, and conserved quantities at spatial infinity}},\ }\href {https://doi.org/10.1063/1.523863} {\bibfield  {journal} {\bibinfo  {journal} {J. Math. Phys.}\ }\textbf {\bibinfo {volume} {19}},\ \bibinfo {pages} {1542} (\bibinfo {year} {1978})}\BibitemShut {NoStop}%
\bibitem [{\citenamefont {Beig}\ and\ \citenamefont {Schmidt}(1982)}]{rbbs}%
  \BibitemOpen
  \bibfield  {author} {\bibinfo {author} {\bibfnamefont {R.}~\bibnamefont {Beig}}\ and\ \bibinfo {author} {\bibfnamefont {B.~G.}\ \bibnamefont {Schmidt}},\ }\bibfield  {title} {\bibinfo {title} {Einstein's equations near spatial infinity},\ }\href@noop {} {\bibfield  {journal} {\bibinfo  {journal} {Communications in Mathematical Physics}\ }\textbf {\bibinfo {volume} {87}},\ \bibinfo {pages} {65} (\bibinfo {year} {1982})}\BibitemShut {NoStop}%
\bibitem [{\citenamefont {Ashtekar}\ and\ \citenamefont {Magnon}(1984)}]{am3+1}%
  \BibitemOpen
  \bibfield  {author} {\bibinfo {author} {\bibfnamefont {A.}~\bibnamefont {Ashtekar}}\ and\ \bibinfo {author} {\bibfnamefont {A.}~\bibnamefont {Magnon}},\ }\bibfield  {title} {\bibinfo {title} {From i° to the 3+1 description of spatial infinity},\ }\href@noop {} {\bibfield  {journal} {\bibinfo  {journal} {Journal of mathematical physics}\ }\textbf {\bibinfo {volume} {25}},\ \bibinfo {pages} {2682} (\bibinfo {year} {1984})}\BibitemShut {NoStop}%
\bibitem [{\citenamefont {Ashtekar}\ and\ \citenamefont {Romano}(1992)}]{aajr}%
  \BibitemOpen
  \bibfield  {author} {\bibinfo {author} {\bibfnamefont {A.}~\bibnamefont {Ashtekar}}\ and\ \bibinfo {author} {\bibfnamefont {J.~D.}\ \bibnamefont {Romano}},\ }\bibfield  {title} {\bibinfo {title} {Spatial infinity as a boundary of spacetime},\ }\href@noop {} {\bibfield  {journal} {\bibinfo  {journal} {Classical and Quantum Gravity}\ }\textbf {\bibinfo {volume} {9}},\ \bibinfo {pages} {1069} (\bibinfo {year} {1992})}\BibitemShut {NoStop}%
\bibitem [{\citenamefont {Henneaux}\ and\ \citenamefont {Troessaert}(2018)}]{henneaux2}%
  \BibitemOpen
  \bibfield  {author} {\bibinfo {author} {\bibfnamefont {M.}~\bibnamefont {Henneaux}}\ and\ \bibinfo {author} {\bibfnamefont {C.}~\bibnamefont {Troessaert}},\ }\bibfield  {title} {\bibinfo {title} {{Hamiltonian structure and asymptotic symmetries of the Einstein-Maxwell system at spatial infinity}},\ }\href@noop {} {\bibfield  {journal} {\bibinfo  {journal} {Journal of High Energy Physics}\ }\textbf {\bibinfo {volume} {2018}},\ \bibinfo {pages} {1} (\bibinfo {year} {2018})}\BibitemShut {NoStop}%
\bibitem [{\citenamefont {Henneaux}\ and\ \citenamefont {Troessaert}(2020)}]{henneaux3}%
  \BibitemOpen
  \bibfield  {author} {\bibinfo {author} {\bibfnamefont {M.}~\bibnamefont {Henneaux}}\ and\ \bibinfo {author} {\bibfnamefont {C.}~\bibnamefont {Troessaert}},\ }\bibfield  {title} {\bibinfo {title} {The asymptotic structure of gravity at spatial infinity in four spacetime dimensions},\ }\href@noop {} {\bibfield  {journal} {\bibinfo  {journal} {Proceedings of the Steklov Institute of Mathematics}\ }\textbf {\bibinfo {volume} {309}},\ \bibinfo {pages} {127} (\bibinfo {year} {2020})}\BibitemShut {NoStop}%
\bibitem [{\citenamefont {Comp\`ere}\ \emph {et~al.}(2023)\citenamefont {Comp\`ere}, \citenamefont {Gralla},\ and\ \citenamefont {Wei}}]{Compere:2023qoa}%
  \BibitemOpen
  \bibfield  {author} {\bibinfo {author} {\bibfnamefont {G.}~\bibnamefont {Comp\`ere}}, \bibinfo {author} {\bibfnamefont {S.~E.}\ \bibnamefont {Gralla}},\ and\ \bibinfo {author} {\bibfnamefont {H.}~\bibnamefont {Wei}},\ }\bibfield  {title} {\bibinfo {title} {{An asymptotic framework for gravitational scattering}},\ }\href {https://doi.org/10.1088/1361-6382/acf5c1} {\bibfield  {journal} {\bibinfo  {journal} {Class. Quant. Grav.}\ }\textbf {\bibinfo {volume} {40}},\ \bibinfo {pages} {205018} (\bibinfo {year} {2023})},\ \Eprint {https://arxiv.org/abs/2303.17124} {arXiv:2303.17124 [gr-qc]} \BibitemShut {NoStop}%
\bibitem [{\citenamefont {Friedrich}(1998)}]{FRIEDRICH199883}%
  \BibitemOpen
  \bibfield  {author} {\bibinfo {author} {\bibfnamefont {H.}~\bibnamefont {Friedrich}},\ }\bibfield  {title} {\bibinfo {title} {Gravitational fields near space-like and null infinity},\ }\href {https://doi.org/https://doi.org/10.1016/S0393-0440(97)82168-7} {\bibfield  {journal} {\bibinfo  {journal} {Journal of Geometry and Physics}\ }\textbf {\bibinfo {volume} {24}},\ \bibinfo {pages} {83} (\bibinfo {year} {1998})}\BibitemShut {NoStop}%
\bibitem [{\citenamefont {Friedrich}(2003)}]{Friedrich:2002ru}%
  \BibitemOpen
  \bibfield  {author} {\bibinfo {author} {\bibfnamefont {H.}~\bibnamefont {Friedrich}},\ }\bibfield  {title} {\bibinfo {title} {{Spin two fields on Minkowski space near space - like and null infinity}},\ }\href {https://doi.org/10.1088/0264-9381/20/1/308} {\bibfield  {journal} {\bibinfo  {journal} {Class. Quant. Grav.}\ }\textbf {\bibinfo {volume} {20}},\ \bibinfo {pages} {101} (\bibinfo {year} {2003})},\ \Eprint {https://arxiv.org/abs/gr-qc/0209034} {arXiv:gr-qc/0209034} \BibitemShut {NoStop}%
\bibitem [{\citenamefont {Fuentealba}\ and\ \citenamefont {Henneaux}(2025)}]{Fuentealba:2024lll}%
  \BibitemOpen
  \bibfield  {author} {\bibinfo {author} {\bibfnamefont {O.}~\bibnamefont {Fuentealba}}\ and\ \bibinfo {author} {\bibfnamefont {M.}~\bibnamefont {Henneaux}},\ }\bibfield  {title} {\bibinfo {title} {{Logarithmic matching between past infinity and future infinity: The massless scalar field in Minkowski space}},\ }\href {https://doi.org/10.1007/JHEP03(2025)081} {\bibfield  {journal} {\bibinfo  {journal} {JHEP}\ }\textbf {\bibinfo {volume} {03}},\ \bibinfo {pages} {081}},\ \Eprint {https://arxiv.org/abs/2412.05088} {arXiv:2412.05088 [gr-qc]} \BibitemShut {NoStop}%
\bibitem [{\citenamefont {Bateman}\ and\ \citenamefont {Project}(1953{\natexlab{a}})}]{bateman_bateman2}%
  \BibitemOpen
  \bibfield  {author} {\bibinfo {author} {\bibfnamefont {H.}~\bibnamefont {Bateman}}\ and\ \bibinfo {author} {\bibfnamefont {B.~M.}\ \bibnamefont {Project}},\ }\href@noop {} {\emph {\bibinfo {title} {{Higher Transcendental Functions. Vol. II}}}}\ (\bibinfo  {publisher} {McGraw-Hill Book Company},\ \bibinfo {year} {1953})\BibitemShut {NoStop}%
\bibitem [{\citenamefont {Geroch}\ \emph {et~al.}(1973)\citenamefont {Geroch}, \citenamefont {Held},\ and\ \citenamefont {Penrose}}]{Geroch:1973am}%
  \BibitemOpen
  \bibfield  {author} {\bibinfo {author} {\bibfnamefont {R.~P.}\ \bibnamefont {Geroch}}, \bibinfo {author} {\bibfnamefont {A.}~\bibnamefont {Held}},\ and\ \bibinfo {author} {\bibfnamefont {R.}~\bibnamefont {Penrose}},\ }\bibfield  {title} {\bibinfo {title} {{A space-time calculus based on pairs of null directions}},\ }\href {https://doi.org/10.1063/1.1666410} {\bibfield  {journal} {\bibinfo  {journal} {J. Math. Phys.}\ }\textbf {\bibinfo {volume} {14}},\ \bibinfo {pages} {874} (\bibinfo {year} {1973})}\BibitemShut {NoStop}%
\bibitem [{\citenamefont {Penrose}\ and\ \citenamefont {Rindler}(2011)}]{Penrose:1985bww}%
  \BibitemOpen
  \bibfield  {author} {\bibinfo {author} {\bibfnamefont {R.}~\bibnamefont {Penrose}}\ and\ \bibinfo {author} {\bibfnamefont {W.}~\bibnamefont {Rindler}},\ }\href {https://doi.org/10.1017/CBO9780511564048} {\emph {\bibinfo {title} {{Spinors and Space-Time}}}},\ Cambridge Monographs on Mathematical Physics\ (\bibinfo  {publisher} {Cambridge Univ. Press},\ \bibinfo {address} {Cambridge, UK},\ \bibinfo {year} {2011})\BibitemShut {NoStop}%
\bibitem [{\citenamefont {Penrose}\ and\ \citenamefont {Rindler}(1988)}]{Penrose:1986ca}%
  \BibitemOpen
  \bibfield  {author} {\bibinfo {author} {\bibfnamefont {R.}~\bibnamefont {Penrose}}\ and\ \bibinfo {author} {\bibfnamefont {W.}~\bibnamefont {Rindler}},\ }\href {https://doi.org/10.1017/CBO9780511524486} {\emph {\bibinfo {title} {{Spinors and Space-Time. Vol. 2: Spinor and Twistor Methods in Space-Time Geometry}}}},\ Cambridge Monographs on Mathematical Physics\ (\bibinfo  {publisher} {Cambridge University Press},\ \bibinfo {year} {1988})\BibitemShut {NoStop}%
\bibitem [{\citenamefont {Price}\ \emph {et~al.}(2007)\citenamefont {Price}, \citenamefont {Shankar},\ and\ \citenamefont {Whiting}}]{Price:2006ke}%
  \BibitemOpen
  \bibfield  {author} {\bibinfo {author} {\bibfnamefont {L.~R.}\ \bibnamefont {Price}}, \bibinfo {author} {\bibfnamefont {K.}~\bibnamefont {Shankar}},\ and\ \bibinfo {author} {\bibfnamefont {B.~F.}\ \bibnamefont {Whiting}},\ }\bibfield  {title} {\bibinfo {title} {{On the existence of radiation gauges in Petrov type II spacetimes}},\ }\href {https://doi.org/10.1088/0264-9381/24/9/014} {\bibfield  {journal} {\bibinfo  {journal} {Class. Quant. Grav.}\ }\textbf {\bibinfo {volume} {24}},\ \bibinfo {pages} {2367} (\bibinfo {year} {2007})},\ \Eprint {https://arxiv.org/abs/gr-qc/0611070} {arXiv:gr-qc/0611070} \BibitemShut {NoStop}%
\bibitem [{\citenamefont {Ashtekar}\ and\ \citenamefont {Magnon-Ashtekar}(1979)}]{Ashtekar:1979xeo}%
  \BibitemOpen
  \bibfield  {author} {\bibinfo {author} {\bibfnamefont {A.}~\bibnamefont {Ashtekar}}\ and\ \bibinfo {author} {\bibfnamefont {A.}~\bibnamefont {Magnon-Ashtekar}},\ }\bibfield  {title} {\bibinfo {title} {{Energy-Momentum in General Relativity}},\ }\href {https://doi.org/10.1103/PhysRevLett.43.181} {\bibfield  {journal} {\bibinfo  {journal} {Phys. Rev. Lett.}\ }\textbf {\bibinfo {volume} {43}},\ \bibinfo {pages} {181} (\bibinfo {year} {1979})}\BibitemShut {NoStop}%
\bibitem [{\citenamefont {Ashtekar}\ and\ \citenamefont {Khera}(2024{\natexlab{a}})}]{Ashtekar:2023wfn}%
  \BibitemOpen
  \bibfield  {author} {\bibinfo {author} {\bibfnamefont {A.}~\bibnamefont {Ashtekar}}\ and\ \bibinfo {author} {\bibfnamefont {N.}~\bibnamefont {Khera}},\ }\bibfield  {title} {\bibinfo {title} {{Unified treatment of null and spatial infinity III: asymptotically minkowski space-times}},\ }\href {https://doi.org/10.1007/JHEP02(2024)210} {\bibfield  {journal} {\bibinfo  {journal} {JHEP}\ }\textbf {\bibinfo {volume} {02}},\ \bibinfo {pages} {210}},\ \Eprint {https://arxiv.org/abs/2311.14130} {arXiv:2311.14130 [gr-qc]} \BibitemShut {NoStop}%
\bibitem [{\citenamefont {Ashtekar}\ and\ \citenamefont {Khera}(2024{\natexlab{b}})}]{Ashtekar:2023zul}%
  \BibitemOpen
  \bibfield  {author} {\bibinfo {author} {\bibfnamefont {A.}~\bibnamefont {Ashtekar}}\ and\ \bibinfo {author} {\bibfnamefont {N.}~\bibnamefont {Khera}},\ }\bibfield  {title} {\bibinfo {title} {{Unified treatment of null and spatial infinity IV: angular momentum at null and spatial infinity}},\ }\href {https://doi.org/10.1007/JHEP01(2024)085} {\bibfield  {journal} {\bibinfo  {journal} {JHEP}\ }\textbf {\bibinfo {volume} {01}},\ \bibinfo {pages} {085}},\ \Eprint {https://arxiv.org/abs/2311.14190} {arXiv:2311.14190 [gr-qc]} \BibitemShut {NoStop}%
\bibitem [{\citenamefont {Prabhu}\ and\ \citenamefont {Shehzad}(2022)}]{kpis}%
  \BibitemOpen
  \bibfield  {author} {\bibinfo {author} {\bibfnamefont {K.}~\bibnamefont {Prabhu}}\ and\ \bibinfo {author} {\bibfnamefont {I.}~\bibnamefont {Shehzad}},\ }\bibfield  {title} {\bibinfo {title} {{Conservation of asymptotic charges from past to future null infinity: Lorentz charges in general relativity}},\ }\href@noop {} {\bibfield  {journal} {\bibinfo  {journal} {Journal of High Energy Physics}\ }\textbf {\bibinfo {volume} {2022}},\ \bibinfo {pages} {1} (\bibinfo {year} {2022})}\BibitemShut {NoStop}%
\bibitem [{\citenamefont {Strominger}(2014)}]{Strominger:2013jfa}%
  \BibitemOpen
  \bibfield  {author} {\bibinfo {author} {\bibfnamefont {A.}~\bibnamefont {Strominger}},\ }\bibfield  {title} {\bibinfo {title} {{On BMS Invariance of Gravitational Scattering}},\ }\href {https://doi.org/10.1007/JHEP07(2014)152} {\bibfield  {journal} {\bibinfo  {journal} {JHEP}\ }\textbf {\bibinfo {volume} {07}},\ \bibinfo {pages} {152}},\ \Eprint {https://arxiv.org/abs/1312.2229} {arXiv:1312.2229 [hep-th]} \BibitemShut {NoStop}%
\bibitem [{\citenamefont {Blanchet}\ \emph {et~al.}(2021)\citenamefont {Blanchet}, \citenamefont {Comp\`ere}, \citenamefont {Faye}, \citenamefont {Oliveri},\ and\ \citenamefont {Seraj}}]{Blanchet:2020ngx}%
  \BibitemOpen
  \bibfield  {author} {\bibinfo {author} {\bibfnamefont {L.}~\bibnamefont {Blanchet}}, \bibinfo {author} {\bibfnamefont {G.}~\bibnamefont {Comp\`ere}}, \bibinfo {author} {\bibfnamefont {G.}~\bibnamefont {Faye}}, \bibinfo {author} {\bibfnamefont {R.}~\bibnamefont {Oliveri}},\ and\ \bibinfo {author} {\bibfnamefont {A.}~\bibnamefont {Seraj}},\ }\bibfield  {title} {\bibinfo {title} {{Multipole expansion of gravitational waves: from harmonic to Bondi coordinates}},\ }\href {https://doi.org/10.1007/JHEP02(2021)029} {\bibfield  {journal} {\bibinfo  {journal} {JHEP}\ }\textbf {\bibinfo {volume} {02}},\ \bibinfo {pages} {029}},\ \Eprint {https://arxiv.org/abs/2011.10000} {arXiv:2011.10000 [gr-qc]} \BibitemShut {NoStop}%
\bibitem [{\citenamefont {Blanchet}\ \emph {et~al.}(2023)\citenamefont {Blanchet}, \citenamefont {Comp\`ere}, \citenamefont {Faye}, \citenamefont {Oliveri},\ and\ \citenamefont {Seraj}}]{Blanchet:2023pce}%
  \BibitemOpen
  \bibfield  {author} {\bibinfo {author} {\bibfnamefont {L.}~\bibnamefont {Blanchet}}, \bibinfo {author} {\bibfnamefont {G.}~\bibnamefont {Comp\`ere}}, \bibinfo {author} {\bibfnamefont {G.}~\bibnamefont {Faye}}, \bibinfo {author} {\bibfnamefont {R.}~\bibnamefont {Oliveri}},\ and\ \bibinfo {author} {\bibfnamefont {A.}~\bibnamefont {Seraj}},\ }\bibfield  {title} {\bibinfo {title} {{Multipole expansion of gravitational waves: memory effects and Bondi aspects}},\ }\href {https://doi.org/10.1007/JHEP07(2023)123} {\bibfield  {journal} {\bibinfo  {journal} {JHEP}\ }\textbf {\bibinfo {volume} {07}},\ \bibinfo {pages} {123}},\ \Eprint {https://arxiv.org/abs/2303.07732} {arXiv:2303.07732 [gr-qc]} \BibitemShut {NoStop}%
\bibitem [{\citenamefont {Laddha}\ and\ \citenamefont {Sen}(2018)}]{Laddha:2018myi}%
  \BibitemOpen
  \bibfield  {author} {\bibinfo {author} {\bibfnamefont {A.}~\bibnamefont {Laddha}}\ and\ \bibinfo {author} {\bibfnamefont {A.}~\bibnamefont {Sen}},\ }\bibfield  {title} {\bibinfo {title} {{Logarithmic Terms in the Soft Expansion in Four Dimensions}},\ }\href {https://doi.org/10.1007/JHEP10(2018)056} {\bibfield  {journal} {\bibinfo  {journal} {JHEP}\ }\textbf {\bibinfo {volume} {10}},\ \bibinfo {pages} {056}},\ \Eprint {https://arxiv.org/abs/1804.09193} {arXiv:1804.09193 [hep-th]} \BibitemShut {NoStop}%
\bibitem [{\citenamefont {Blanchet}(2014)}]{Blanchet:2013haa}%
  \BibitemOpen
  \bibfield  {author} {\bibinfo {author} {\bibfnamefont {L.}~\bibnamefont {Blanchet}},\ }\bibfield  {title} {\bibinfo {title} {{Post-Newtonian Theory for Gravitational Waves}},\ }\href {https://doi.org/10.12942/lrr-2014-2} {\bibfield  {journal} {\bibinfo  {journal} {Living Rev. Rel.}\ }\textbf {\bibinfo {volume} {17}},\ \bibinfo {pages} {2} (\bibinfo {year} {2014})},\ \Eprint {https://arxiv.org/abs/1310.1528} {arXiv:1310.1528 [gr-qc]} \BibitemShut {NoStop}%
\bibitem [{\citenamefont {{\'A}lvares}\ and\ \citenamefont {Va{\~n}o-Vin{\~{u}}ales}(2025)}]{Alvares:2025pbi}%
  \BibitemOpen
  \bibfield  {author} {\bibinfo {author} {\bibfnamefont {J.~D.}\ \bibnamefont {{\'A}lvares}}\ and\ \bibinfo {author} {\bibfnamefont {A.}~\bibnamefont {Va{\~n}o-Vin{\~{u}}ales}},\ }\bibfield  {title} {\bibinfo {title} {{Charged Scalar Field at Future Null Infinity via Nonlinear Hyperboloidal Evolution}},\ }\href@noop {} {\  (\bibinfo {year} {2025})},\ \Eprint {https://arxiv.org/abs/2506.15311} {arXiv:2506.15311 [gr-qc]} \BibitemShut {NoStop}%
\bibitem [{\citenamefont {Bateman}\ and\ \citenamefont {Project}(1953{\natexlab{b}})}]{bateman_bateman}%
  \BibitemOpen
  \bibfield  {author} {\bibinfo {author} {\bibfnamefont {H.}~\bibnamefont {Bateman}}\ and\ \bibinfo {author} {\bibfnamefont {B.~M.}\ \bibnamefont {Project}},\ }\href@noop {} {\emph {\bibinfo {title} {{Higher Transcendental Functions. Vol. I}}}}\ (\bibinfo  {publisher} {McGraw-Hill Book Company},\ \bibinfo {year} {1953})\BibitemShut {NoStop}%
\end{thebibliography}%

\end{document}